\documentclass[preprint,12pt]{elsarticle}
\usepackage[utf8]{inputenc}
\usepackage[T1]{fontenc}
\usepackage{indentfirst}
\usepackage{graphicx}
\usepackage{amsmath}
\usepackage{appendix}
\usepackage{ccaption}
\usepackage{vmargin}
\usepackage{textcomp}
\usepackage{float}

\setmarginsrb{3cm}{2cm}{2cm}{2cm}{2ex}{2ex}{2ex}{3ex}

\usepackage{amssymb}

\begin{document}

\begin{frontmatter}

\title{Simulation of radio emission from cosmic ray air shower with SELFAS2}

\author{Vincent Marin}
\author{Beno\^it Revenu}

\address{SUBATECH, CNRS-IN2P3 4 rue Alfred Kastler BP 20722 44307 NANTES CEDEX 03 France}

\begin{abstract}

We present a microscopic computation of the radio emission from air showers initiated by ultra-high energy cosmic rays in the atmosphere. The strategy adopted is to compute each secondary particle contribution of the electromagnetic component and to construct the total signal at any location. SELFAS2 is a code which doesn't rely on air shower generators like AIRES or CORSIKA and it is based on the concept of air shower universality which makes it completely autonomous. Each positron and electron of the air shower is generated randomly following relevant distributions and tracking them along their travel in the atmosphere. We confirm in this paper earlier results that the radio emission is mainly due to the time derivative of the transverse current and the time derivative of the charge excess. The time derivative of the transverse current created by systematic deviations of charges in the geomagnetic field is usually dominant compared to the charge excess contribution except for the case of an air shower parallel to the geomagnetic field.

\end{abstract}

\begin{keyword}
cosmic rays, air shower, radio emission, radiation of moving charges.
\end{keyword}

\end{frontmatter}

\section{Introduction}
In 1965, Jelley \textit{et al.} \cite{key1} and Allan in 1971 \cite{key2}  established that extensive air showers (EAS) induced by ultra high energy cosmic ray emit a brief radio emission in the range of few hundreds of MHz during its development in the atmosphere. In the last years, progress made in electronics have permitted the realization of extremely fast radio-detectors used in the CODALEMA \cite{key3}, LOPES \cite{key3b} and AERA \cite{key7} experiments. It is clear since \cite{key2,key31,key32} and recently \cite{key4,key3} that the Lorentz force induced by the geomagnetic field acting on positrons and electrons of the shower is the dominant source of the EAS radio emission. The consequences of this effect are visible through a strong asymmetry in the counting rate as a function of arrival directions (see for instance \cite{key3,key3b,key9}). Recent experimental analysis shows that the lateral profile of the electric field deposited by an EAS on the ground permits to give a good estimation of the primary energy \cite{key111,key11}. 

In parallel to these experimental results, there are also various theoretical approaches, from simple fast semi-analytical models (see Scholten \textit{et al.} \cite{key16bis}, Chauvin et al. \cite{key12}) to more detailed models (see Huege \textit{et al} \cite{key02}). Initially based on the synchrotron radiation proposed by Falcke and Gorham \cite{key13}, the electric field in REAS is now computed using the end point formalism \cite{key25} applied on each secondary electrons and positrons of the air shower. The total electric field at a given observation point is obtained summing all electric field contributions coming from the shower. The geometry of the shower (pancake thickness, lateral extension) is then taken into account. This characteristic appears to be important if we want to compute the electric field close to the shower axis. In MGMR \cite{key16,key16bis} (Scholten \textit{et al.}), the total electric field is sourced by the global macroscopic transverse current due to the systematic deviation of electrons and positrons in the geomagnetic field. Maxwell equations are applied to this global current concentrated along the air shower axis, taking into consideration the thickness of the air shower pancake. Due to the secondary electrons excess during the air shower development (mainly due to positrons annihilation and knock-out electrons during the air shower development in the atmosphere), the residual negative charge variation induces a second contribution to the total EAS radio emission \cite{key01,key001,key02,key19}. This second contribution (charge excess contribution) is not EAS arrival direction dependent while it is the case for the contribution due to systematic deviation of particles in the geomagnetic field (geomagnetic contribution). This characteristics implies that the fraction of charge excess contribution to the total EAS radio signal also depends on the arrival direction. In most cases, the geomagnetic contribution is dominant, but for arrival directions close to the magnetic field orientation, the Lorentz force due to the geomagnetic field vanishes and the charge excess contribution can become dominant in the total EAS radio signal.

Our approach is based on a microscopic description of the shower using the concept of "age" and "shower universality" first proposed in \cite{key333} to study the longitudinal development of purely electromagnetic showers. The use of the relevant distributions for EAS secondary electrons and positrons extracted from \cite{key271,key272,key27} (longitudinal profile, particle energy, vertical and horizontal momentum angle, lateral distance, and time distribution of the shower front), permits to avoid the heavy use of EAS generators to generate air showers in SELFAS2 and makes the simulation completely autonomous. Thanks to these distributions implemented in SELFAS2, no large simplifications are made on the characteristics of the electromagnetic shower component. The particles generated in SELFAS2 by Monte Carlo simulation, are tracked along their trajectory to compute their individual electric field contribution to the total electric field emitted by the air shower. With this approach, the characteristics of the evolutive spatial density of charge (emissive area) in the shower and the systematic drift of electrons and positrons due to the geomagnetic field are then naturally taken into account in SELFAS2. 

In section 2, we develop the formalism adopted in SELFAS2 to compute the electric field emitted by a single charge with a finite life time which undergoes an acceleration due to the presence of a magnetic field. We will describe the strategy adopted to sum correctly the contributions given by each particle during its existence in the air shower. In section 3 we describe how the electromagnetic component EAS is generated to give to each particle initial conditions for their travel in the atmosphere. Section 4 is devoted to a discussion from a first example of a vertical $10^{17}$ eV air shower.

\section{Theory: Maxwell equations for moving charges}
\subsection{Flash back on modelings}
Historically, many modelings of EAS radio emission were based on the well known equation of the electric field emitted by a relativistic moving charge which undergoes an acceleration. This equation is obtained solving the Maxwell equation for a retarded time using retarded potential (see for example \cite{key20}). The resulting electric field of a charge moving with a velocity $\boldsymbol{\beta}$ in the lab frame is given by:
\begin{equation}
\boldsymbol{E}(\boldsymbol{x},t)=\frac{q}{4 \pi  \epsilon _0}\left[\frac{\boldsymbol{n}-\boldsymbol{\beta }}{\gamma ^2\left(1-\boldsymbol{\beta }.\boldsymbol{n}\right)^3R^2}\right]_{ret}+\frac{q}{4 \pi  \epsilon _0 c}\left[\frac{\boldsymbol{n}\times \left\{\left(\boldsymbol{n}-\boldsymbol{\beta }\right)\times \boldsymbol{\dot{\beta }}\right\}}{\left(1-\boldsymbol{\beta }.\boldsymbol{n}\right)^{3}R}\right]_{ret}
\label{EqJack} 
\end{equation} 
which is obtained at the observation point $\boldsymbol{x}$ at the time $t=t_{ret}+R(t_{ret})/c$ with $\boldsymbol{n}=(\boldsymbol{x}-\boldsymbol{r}(t_{ret}))/R(t_{ret})$ and $R=|\boldsymbol{x}-\boldsymbol{r}(t_{ret})|$. Two terms appear in this equation which correspond to a coulombian contribution for the first one and to a radiative contribution for the second one due to the acceleration $\dot{\boldsymbol{\beta}}$ of the source. 

In many approaches (\cite{key21,key22,key23,key12,key24}), the total electric field generated by the EAS is obtained after summation of the individual electric field performed using Eq.\ref{EqJack} or simply replacing $q$ in Eq.\ref{EqJack} by the global varying charge $Q(t)$ of the EAS. But, as it was recently discussed in \cite{key19} this way to describe the present problem is not correct because Eq.\ref{EqJack} is obtained for a charge $q$ which is not time dependent. To overcome this problem, a solution has been proposed in \cite{key02,key25}. The solution proposed in this paper and adopted in SELFAS2 is different. 

\subsection{Electric field of a point like source with a finite life time}

Starting from Maxwell equations and using the wave equation for the scalar potential $\Phi$ and the vector potential $\boldsymbol{A}$ (see for instance \cite{key20}) we have:
\begin{eqnarray}
\nabla ^2\Phi -\frac{1}{c^2}\frac{\partial^2 \Phi }{\partial^2 t}&=&-\frac{\rho }{\epsilon _0}\\
\nabla ^2\boldsymbol{A} -\frac{1}{c^2}\frac{\partial^2 \boldsymbol{A} }{\partial^2 t}&=&-\mu _0 \boldsymbol{J}
\end{eqnarray}
where $\rho$ is the charge and $\boldsymbol{J}$ the current densities. Using the expression of the electric field:
\begin{eqnarray}
\boldsymbol{E}&=&-\boldsymbol{\nabla}\Phi-\frac{\partial \boldsymbol{A}}{\partial t}
\label{EqMaxwell} 
\end{eqnarray}
we can express the electric field $\boldsymbol{E}$ as a function of the charge $\rho$ and the current densities $\boldsymbol{J}$:
\begin{eqnarray}
\nabla ^2\boldsymbol{E} -\frac{1}{c^2}\frac{\partial^2 \boldsymbol{E} }{\partial^2 t}&=&-\frac{1 }{\epsilon _0}\left(-\boldsymbol{\nabla}\rho-\frac{1}{c^2}\frac{\partial \boldsymbol{J}}{\partial t}\right)
\end{eqnarray}
This differential equation can be solved using a retarded solution (Green function). It gives: 
\begin{eqnarray}        
\boldsymbol{E}(\boldsymbol{x},t)&=&\frac{1}{4\pi  \epsilon _0}\int d^3x'dt'\frac{1}{R}\left[-\boldsymbol{\nabla}'\rho-\frac{1}{c^2}\frac{\partial \boldsymbol{J}}{\partial t'}\right]_{\text{ret}}\delta \left\{t'-\left(t-\frac{|\boldsymbol{x}-\boldsymbol{x}'|}{c}\right)\right\}
\label{E}
\end{eqnarray}
where we can define $\boldsymbol{R}=\boldsymbol{x}-\boldsymbol{x}'$ and $R=|\boldsymbol{R}|$ with $\boldsymbol{x}'$ the source position at the retarded time $t'$ and $\boldsymbol{x}$ the obervation point. Here, $\boldsymbol{\nabla}'$ and $\frac{\partial }{\partial t'}$ must be considered at retarded time. This remark is important, because it is also the case for the expression of the field in Eq.\ref{EqMaxwell} which should be considered at the retarded time; the variable $t$ in Eq.\ref{EqMaxwell} is the time at the instant of emission. A particular treatment must be done in order to get $\boldsymbol{\nabla}$ and $\frac{\partial }{\partial t}$ out from the retarded brackets in order to obtain a final expression of $\boldsymbol{E}$ as a function of $t$ and not $t'$.
Due to the dependance of $t'$ on $\boldsymbol{x'}$ given by the relation between $t'$ and $t$:
\begin{eqnarray}
t=t'+\frac{R}{c}=t'+\frac{|\boldsymbol{x}-\boldsymbol{x}'|}{c}
\end{eqnarray}
the expression $\left[\boldsymbol{\nabla}'\rho\right]_{\text{ret}}$ can be transformed in (see \cite{key20}):
\begin{eqnarray}
\left[\boldsymbol{\nabla}'\rho\right]_{\text{ret}}&=&\boldsymbol{\nabla}'\left[\rho\right]_{\text{ret}}-\frac{\boldsymbol{n}}{c}\left[\frac{\partial \rho}{\partial t'}\right]_{\text{ret}}
\end{eqnarray}
where $\boldsymbol{n}=\boldsymbol{R}/R$ is the unit vector between the observation position and the source, oriented toward the observation position. Then Eq.\ref{E} becomes:
\begin{eqnarray}
\begin{split}
\boldsymbol{E}(\boldsymbol{x},t)=\frac{1}{4\pi  \epsilon _0}\int d^3x'dt'\Bigg\{-&\frac{1}{R}\boldsymbol{\nabla}'\left[\rho(\boldsymbol{x}',t')\right]_{\text{ret}}+\frac{\boldsymbol{n}}{cR}\left[\frac{\partial \rho(\boldsymbol{x}',t')}{\partial t'}\right]_{\text{ret}}\\
&-\frac{1}{c^2R}\left[\frac{\partial \boldsymbol{J}(\boldsymbol{x}',t')}{\partial t'}\right]_{\text{ret}}\Bigg\}  \delta \left\{t'-\left(t-\frac{|\boldsymbol{x}-\boldsymbol{x}'|}{c}\right)\right\}
\end{split}
\end{eqnarray}
Using the fact that the charge distribution is spatially localized on a point, the first term in this expression can be rewritten:
\begin{eqnarray}
\int d^3x'\frac{1}{R}\boldsymbol{\nabla}'\left[\rho(\boldsymbol{x}',t')\right]_{\text{ret}}=-\int d^3x'\frac{\boldsymbol{n}}{R^2}\left[\rho(\boldsymbol{x}',t')\right]_{\text{ret}}
\end{eqnarray}
We can finally express the electric field as:
\begin{eqnarray}
\begin{split}
\boldsymbol{E}(\boldsymbol{x},t)=\frac{1}{4\pi  \epsilon _0}\int d^3x'dt'\Bigg\{\frac{\boldsymbol{n}}{R^2}&\left[\rho(\boldsymbol{x}',t')\right]_{\text{ret}}+\frac{\boldsymbol{n}}{cR}\left[\frac{\partial \rho(\boldsymbol{x}',t')}{\partial t'}\right]_{\text{ret}}\\
&-\frac{1}{c^2R}\left[\frac{\partial \boldsymbol{J}(\boldsymbol{x}',t')}{\partial t'}\right]_{\text{ret}}\Bigg\}  \delta \left\{t'-\left(t-\frac{|\boldsymbol{x}-\boldsymbol{x}'|}{c}\right)\right\}
\end{split}
\label{Jefimenko}
\end{eqnarray}
which is known as the Jefimenko's generalization of the coulomb law (see \cite{key20} fore more details). The idea is now to express this formula in the case of a moving particle with a finite life time. The expressions for the charge and the current are:
\begin{eqnarray}
\rho(\boldsymbol{x}',t')&=&q [\theta (t'-t_1)-\theta (t'-t_2)] \delta ^3(\boldsymbol{x}'-\boldsymbol{x}_0(t')) 
\label{rho}
\end{eqnarray}
\begin{eqnarray}
\boldsymbol{J}(\boldsymbol{x}',t')&=&\rho(\boldsymbol{x}',t') \boldsymbol{v}(t') 
\label{J}
\end{eqnarray}
where the retarded instant $t_1$ corresponds to the creation of the moving charge (particle) by sudden acceleration from the state of rest to $v$ and where the retarded instant $t_2$ corresponds to the cancellation of the charge by sudden deceleration from $v$ to the state of rest.

Injecting Eq.\ref{rho} and Eq.\ref{J} in Eq.\ref{Jefimenko} and using the fact that $R$ doesn't explicitly depends on $t$ we perform the integrations over space and time. We obtain: 
\begin{eqnarray}
\begin{split}
\boldsymbol{E}(\boldsymbol{x},t)=\frac{1}{4\pi  \epsilon _0}&\Bigg\{\left[\frac{\boldsymbol{n}q(t_{ret})}{R^2(1-\boldsymbol{\beta}.\boldsymbol{n})}\right]_{\text{ret}} +  \frac{1}{c}\frac{\partial}{\partial t}\left[\frac{\boldsymbol{n}q(t_{ret})}{R(1-\boldsymbol{\beta}.\boldsymbol{n})}\right]_{\text{ret}} -   \frac{1}{c^2}\frac{\partial}{\partial t}\left[\frac{\boldsymbol{v}q(t_{ret})}{R(1-\boldsymbol{\beta}.\boldsymbol{n})}\right]_{\text{ret}}\Bigg\}
\end{split}
\label{EField}
\end{eqnarray}
with
\begin{eqnarray}
q(t_{ret})=q [\theta (t_{ret}-t_1)-\theta (t_{ret}-t_2)]
\label{charge}
\end{eqnarray}

The electric field emitted by a moving charge with finite life time is the summation of three contributions. The first one is a static contribution directly linked to the scalar potential. It appears by the simple fact that the particle exists. The second term, which is probably the less evident term to guess when we see Eq.\ref{EqMaxwell}, comes from the transformation of the retarded scalar potential gradient. It corresponds to the direct charge variation contribution. The third one is directly the time derivative of the current. For the case of an isolated particle, it corresponds to its radiation emitted if its undergoes an acceleration, like a deflection for instance. For a particle with a finite lifetime, as described by Eq.\ref{charge}, the charge variation appears only at the start point and the end point of the particle trajectory. Due to the description of the charge existence with a Heaviside-step function in Eq.\ref{charge}, the instantaneous charge variation at $t_1$ and $t_2$ will give un-physical divergences for the calculation of the instantaneous electric field at these instants. However, these un-physical sharp pulses at $t_1$ and $t_2$ can be ignored due to the fact that they give electric field contributions out of the frequency range of interest (in our problem, we focus on electric field below 500 MHz). We will see in the next section that for an ensemble of charge, the collective effect will give a charge variation which, now, has a contribution in our frequency range.

\subsection{Electric field of an ensemble of charges} 

In the air shower, particles are permanently created and annihilated following the distribution of particles number during the air shower development. We compute the contribution of each particle during its life time in the shower and we do the summation at a given observation point.

The total electric field created by an ensemble of charges observed at the position $\boldsymbol{x}$ is given by:
\begin{eqnarray}
\boldsymbol{E}_{tot}(\boldsymbol{x},t)=\sum _{i=1} ^{\zeta}\boldsymbol{E_i}(\boldsymbol{x},t)
\end{eqnarray}
where $\zeta$ is the number of particle with $\boldsymbol{r}(t_{ret})$ and $t_{ret}$ verifying $t=t_{ret}+|\boldsymbol{x}-\boldsymbol{r}(t_{ret})|/c$. Performing the time derivative of each particle contribution in Eq.\ref{EField} before doing the summation is not the good strategy. The formulation of Eq.\ref{EField} permits us, by a simple law of derivation composition, to move the summation symbol inside the time derivative operator:
\begin{eqnarray}
\begin{split}
\boldsymbol{E}_{tot}(\boldsymbol{x},t)=\frac{1}{4\pi  \epsilon _0}\Bigg\{\sum _{i=1} ^{\zeta}\left[\frac{\boldsymbol{n_i}q_i(t_{ret})}{R_i^2(1-\boldsymbol{\beta}_i.\boldsymbol{n}_i)}\right]_{\text{ret}} + & \frac{1}{c}\frac{\partial}{\partial t}\sum _{i=1} ^{\zeta}\left[\frac{\boldsymbol{n}_iq_i(t_{ret})}{R_i(1-\boldsymbol{\beta}_i.\boldsymbol{n}_i)}\right]_{\text{ret}}\\
&-\frac{1}{c^2}\frac{\partial}{\partial t}\sum _{i=1} ^{\zeta}\left[\frac{\boldsymbol{v}_iq_i(t_{ret})}{R_i(1-\boldsymbol{\beta}_i.\boldsymbol{n}_i)}\right]_{\text{ret}}\Bigg\}
\end{split}
\label{SumField}
\end{eqnarray}

This equation summarizes the strategy adopted in SELFAS to compute the electric field at any observation point. The strategy will be to construct separately the three parts of this equation and to realize the time derivative operation after all particles have been considered. The time derivative of charge makes now perfect sense in our frequency domain due to the collective effect. The macroscopic charge variation induced by the evolution of the number of particles along the air shower development will give a low frequency contribution to the electric field.

To summarize, we can finally decompose the EAS radio emission process into various contributions:
\begin{itemize}
\item the static contribution of all particles (term $\boldsymbol{St}$ in the following equation);
\item the time derivative of the net charge excess (term $\boldsymbol{C}$ in the following equation);
\item the time derivative of the transverse current due to systematic separation in the geomagnetic field of electrons and positrons from the charge excess and from the charge symmetry: electrons and positrons in equal quantity in opposition to charge excess (term $\boldsymbol{Cu}$ in the following equation).
\end{itemize}

For convenience, we define: 
\begin{eqnarray}
\boldsymbol{st}_{i}(\boldsymbol{x},t)=\left[\frac{\boldsymbol{n_i}q_i(t_{ret})}{R_i^2(1-\boldsymbol{\beta}_i.\boldsymbol{n}_i)}\right]_{\text{ret}}  \hspace{3ex} \text{and} \hspace{3ex} \boldsymbol{St}(\boldsymbol{x},t)=\sum _{i=1} ^{\zeta}\boldsymbol{st}_{i}(\boldsymbol{x},t) 
\label{eq19}
\end{eqnarray}
\begin{eqnarray}
\boldsymbol{c}_{i}(\boldsymbol{x},t)=\left[\frac{\boldsymbol{n}_iq_i(t_{ret})}{R_i(1-\boldsymbol{\beta}_i.\boldsymbol{n}_i)}\right]_{\text{ret}} \hspace{3ex} \text{and} \hspace{3ex} \boldsymbol{C}(\boldsymbol{x},t)=\sum _{i=1} ^{\zeta}\boldsymbol{c}_{i}(\boldsymbol{x},t) 
\label{eq20}
\end{eqnarray}
\begin{eqnarray}
\boldsymbol{cu}_{i}(\boldsymbol{x},t)=\left[\frac{\boldsymbol{v}_iq_i(t_{ret})}{R_i(1-\boldsymbol{\beta}_i.\boldsymbol{n}_i)}\right]_{\text{ret}} \hspace{3ex} \text{and} \hspace{3ex} \boldsymbol{Cu}(\boldsymbol{x},t)=\sum _{i=1} ^{\zeta}\boldsymbol{cu}_{i}(\boldsymbol{x},t) 
\label{eq21}
\end{eqnarray}

Where $\boldsymbol{st}(\boldsymbol{x},t)$, $\boldsymbol{c}(\boldsymbol{x},t)$, $\boldsymbol{cu}(\boldsymbol{x},t)$ are the contributions of the $i^{th}$ particle to $\boldsymbol{St}(\boldsymbol{x},t)$, $\boldsymbol{C}(\boldsymbol{x},t)$, $\boldsymbol{Cu}(\boldsymbol{x},t)$ which are the total static contribution, the total charge contribution and  the total current contribution. The total field is then:
\begin{eqnarray}
\boldsymbol{E}_{tot}(\boldsymbol{x},t)=\frac{1}{4\pi  \epsilon _0} \left( \boldsymbol{St}(\boldsymbol{x},t)+  \frac{1}{c}\frac{\partial}{\partial t}\boldsymbol{C}(\boldsymbol{x},t) - \frac{1}{c^2}\frac{\partial}{\partial t}\boldsymbol{Cu}(\boldsymbol{x},t)\right).
\label{SumFiel}
\end{eqnarray}

\section{SELFAS2 algorithm, air shower generation}
\subsection{Air shower generation, longitudinal profile, universality}
What we need is to track each particle generated in SELFAS2 along its trajectory to compute the electric field as a function of time. To save a lot of computation time, only a fraction of the real particle number is generated. The statistical weight and the trajectory length are the same in SELFAS2 for each particle generated. The statistical weight is calculated by a simple proportionality law between the particle number generated in the simulation, and the real particle number which should be generated to obtain the complete air shower requested. To respect the distribution of the instantaneous number of particles in the shower $N(X)$ (where $X$ is the depth crossed by the shower), we can easily show that the random injection depth given to each particle can be performed by Monte Carlo simulation following :
\begin{eqnarray}
I_p(X)\approx\frac{N(X+l)}{l}
\end{eqnarray}
where $l$,  the length of the individual particle track, is sufficiently small with respect to the shower length. 

In SELFAS2, the evolution of the particle number is given by the Greisen-Iljina-Linsley (GIL) parameterization \cite{key33} based on the Greisen relations \cite{key031} and on a variant for describing nucleus-initiated showers \cite{key32}. This parameterization gives the number of electrons and positrons as a function of the primary energy ($E_p$) and its nature (or mass $A$):
\begin{eqnarray}
N(E_p,A,t)=\frac{E_p}{E_l}e^{t-t_{max}-2\ln(s)}
\end{eqnarray}
where 
\begin{eqnarray}
t=\frac{X-X_1}{X_0} \hspace{0.5cm} , \hspace{0.5cm} t_{max}=a+b\,\left(\ln\frac{E_p}{Ec}-\ln A\right) \hspace{0.5cm} \text{and} \hspace{0.5cm} s=\frac{2t}{t+t_{max}}
\end{eqnarray} 
with $E_l=1450$ MeV is a normalization factor, $E_c=81$ MeV is the critical energy from Greisen formula, $a=1.7$ is an offset parameter, $b=0.76$ is the value of the elongation rate coming from adjusted data. $X$ here is the atmospheric depth measured from the first interaction point $X_1$ in g\,cm$^{-2}$ and $X_0=36.7$ g\,cm$^{-2}$ is the radiation length of electrons in air. The first interaction length $X_1$ can be fixed in SELFAS2 or can be chosen randomly following the cross section of the primary with air \cite{key34}. 

Using these parameterizations, the evolution of the number of electrons and positrons is then completely integrated in SELFAS2, and is completely defined with the energy and the nature of the primary fixed in the input file by the user. 

\subsection{Particles initial conditions}
With a large set of EAS simulated with CORSIKA, Lafèbre \textit{et al} \cite{key27} provide multi-dimensional parameterizations for the electron-positron distributions in terms of particle energy, vertical and horizontal momentum angle, lateral distance, and time distribution of the shower front. They parameterize all these distributions as a function of the relative evolution time, $t$ (defined differently as in previous section for the GIL parameterization):
\begin{eqnarray}
t\equiv \frac{X-X_{max}}{X_0}
\end{eqnarray}
\begin{figure}
\begin{center}
\includegraphics[scale=0.582]{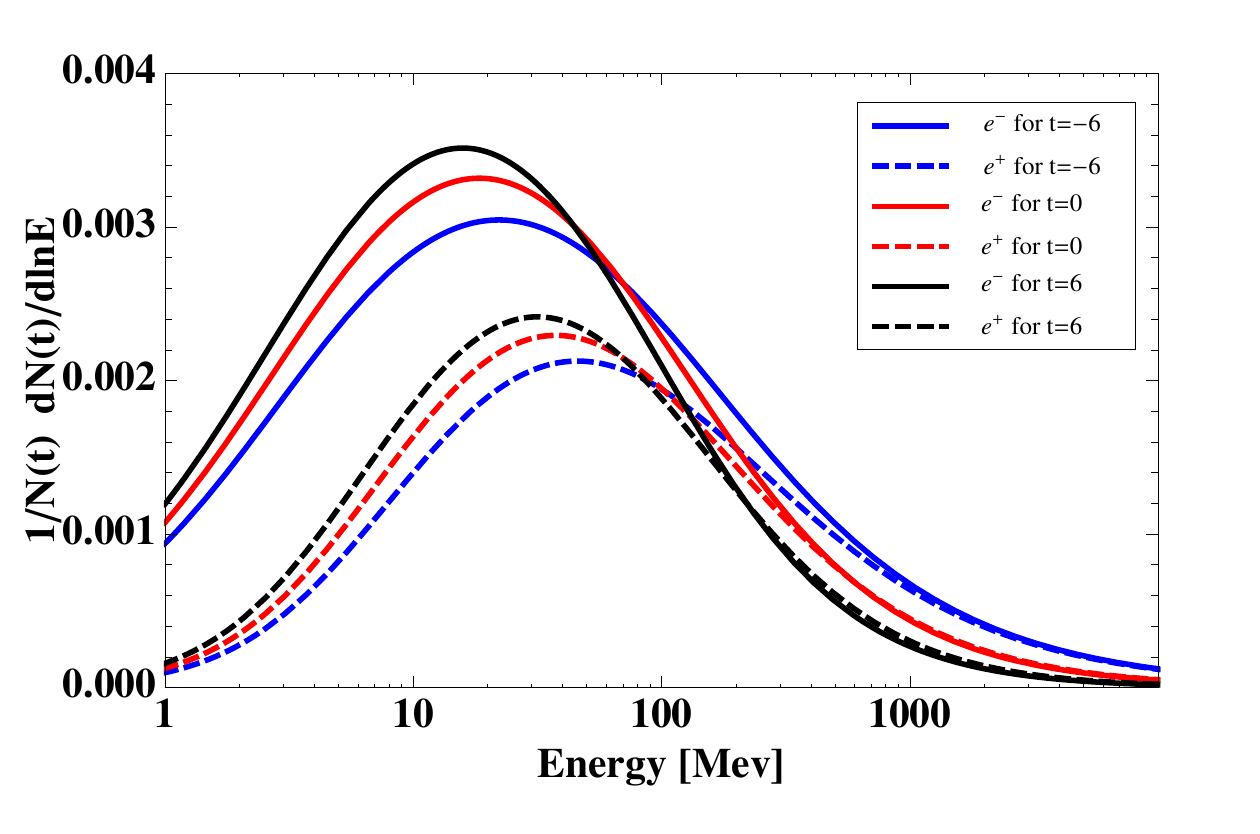}
\includegraphics[scale=0.582]{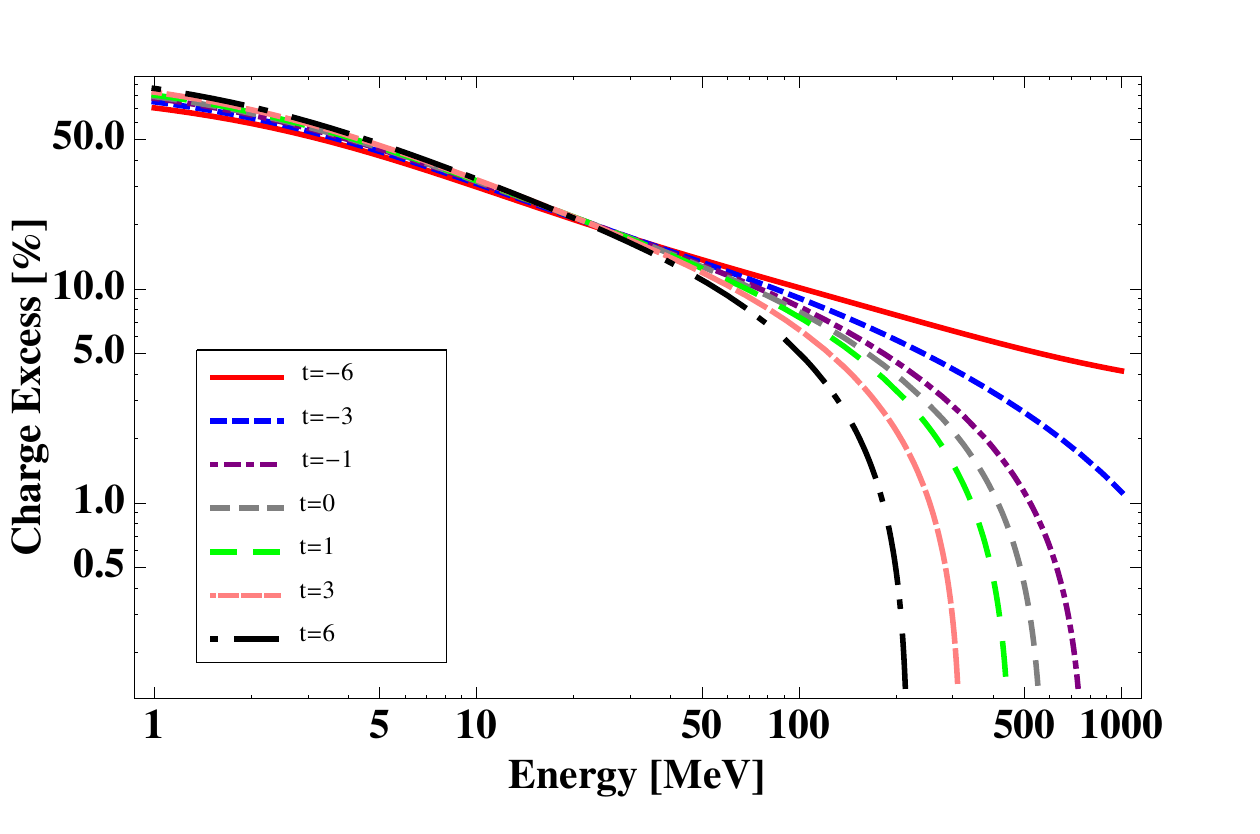}
\end{center}
\caption{\footnotesize{Left: Energy distributions of electrons and positrons in EAS for different evolution stages. Right: Charge excess characterization as a function of particle energies for different evolution stages of the EAS. The complete set of parameterizations for particle energy distributions, vertical and horizontal momentum angle distribution, lateral distance distributions, and time distributions of the shower front is available in \cite{key27}.}}
\label{Energy}
\end{figure}
where $X$ is the integrated thickness of the atmosphere crossed, $X_0=36.7$ g$\,$cm$^{-2}$ is the radiation length of electrons in air and $X_{max}$ is the atmospheric depth for which the number of particles in the air shower is the maximum. As an example, the evolution of the energy distribution for electrons and positrons is shown in Fig.\ref{Energy} left for different relative evolution stages. A parameterization of the charge excess is also available in \cite{key27} as a function of the particle energy and the relative evolution stages of the EAS. We show some results in Fig.\ref{Energy} right, for different evolution stages. We see that the charge excess decreases significantly at high energy (above 100~MeV) during the EAS development. When an electron is generated in SELFAS2 with a random energy $\epsilon$, at a random evolution stage $t$, the corresponding charge excess fraction is calculated. Then, with a simple random law, the electron is considered coming from the charge excess or not. 

All the geometrical parameterizations described in \cite{key27} are implemented in SELFAS2. At each particle generation, SELFAS2 gives randomly a set of initial conditions following all these distributions. The universality of the analytical geometrical descriptions permits to generate air showers in the range of $10^{17}$ to $10^{20}$ eV with a good accuracy for showers coming with zenith angles from 0° to 60°. Only the general characteristics of the required air shower is needed by SELFAS2: primary energy, azimuthal and zenith angles. The geographical characteristics of the site where the simulation must be realized have to be specified as well: orientation of the geomagnetic field, its intensity and the ground altitude (the altitude zero corresponds to the sea level). 

\subsection{Particles propagation and electric field computation}
In order to compute energy losses and multiple scattering of electrons and positrons in the atmosphere, the length $l$, of an individual particle track is divided in short tracks corresponding to a depth fixed to $X_{\text{track}}^{\text{short}}$ (the choice of the resolution is discussed in the next section 4). Between two consecutive points, each particle undergoes a deviation due to the geomagnetic field and multiple scattering. These two processes are computed independently and the final new position in space after one short track results from the combination of the two deviations. The geomagnetic field deviation is calculated solving the equation of motion and the deviation from multiple scattering is performed by Monte-Carlo simulation following the theory of multiple scattering through small angles \cite{key30}. 

\begin{figure}
\includegraphics[scale=0.45]{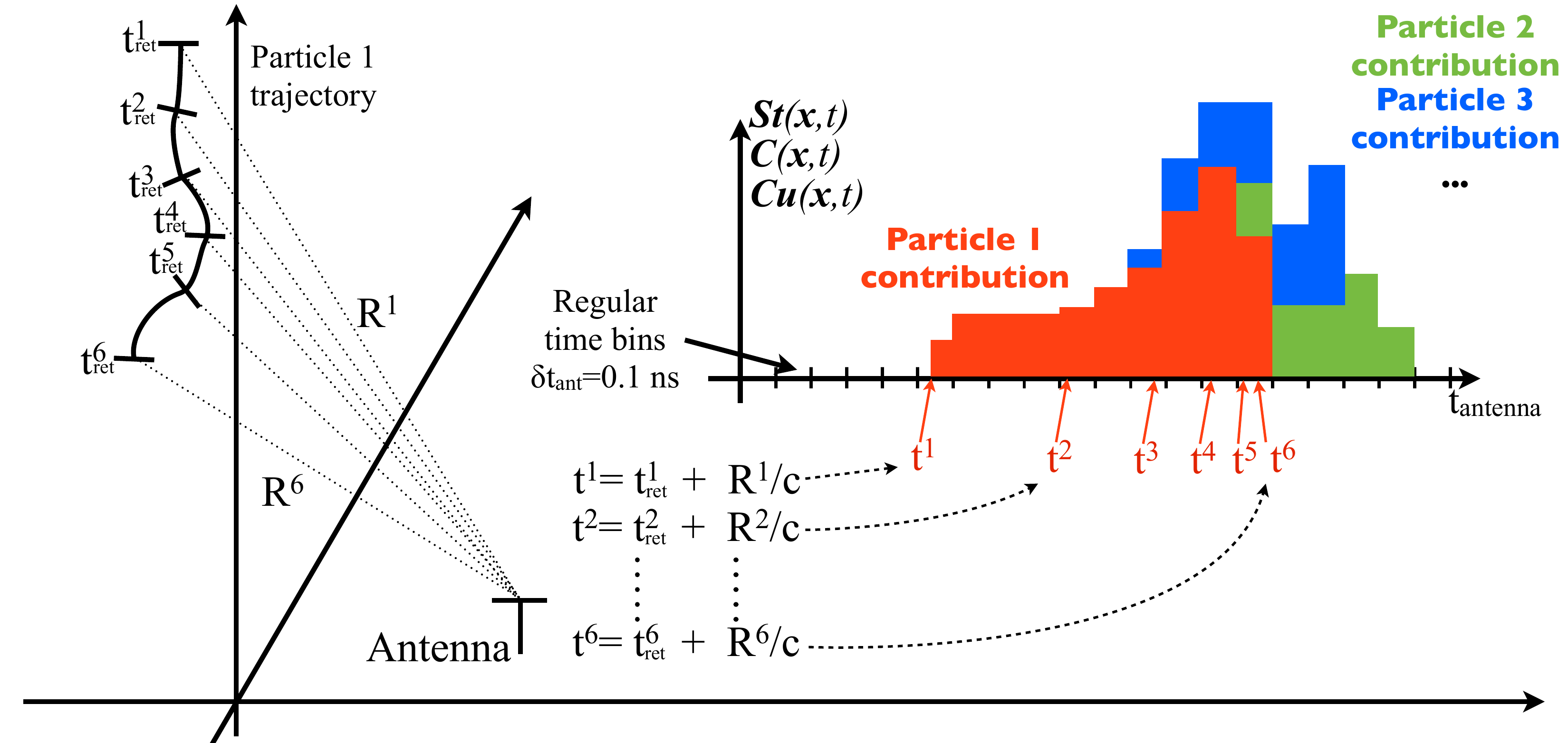}
\caption{\footnotesize{The length of an individual particle track of the particle trajectory is divided in short tracks allowing us to compute energy losses and deviations due to geomagnetic field and to multiple scattering. At the starting point and at each end of short track, $\boldsymbol{st}_{i}(\boldsymbol{x},t)$, $\boldsymbol{c}_{i}(\boldsymbol{x},t)$ and $\boldsymbol{cu}_{i}(\boldsymbol{x},t)$ are calculated independently and added to three independent histograms corresponding to $\boldsymbol{St}(\boldsymbol{x},t)$, $\boldsymbol{C}(\boldsymbol{x},t)$ and $\boldsymbol{Cu}(\boldsymbol{x},t)$. Once all particles have been considered, the time derivatives for $\boldsymbol{C}(\boldsymbol{x},t)$ and $\boldsymbol{Cu}(\boldsymbol{x},t)$ are performed and the complete electric field at this position is obtained summing up the three total contributions.}}
\label{Picture}
\end{figure}

To compute the electric field contribution along each particle trajectory, the position and the velocity of the particle are calculated every $X_{\text{track}}^{\text{short}}$, so we are able to compute the instantaneous $\boldsymbol{st}_{i}(x,t)$, $\boldsymbol{c}_{i}(x,t)$, and $\boldsymbol{cu}_{i}(x,t)$ along the particle trajectory, received at the time $t=t_{\text{ret}}+R/c$ at the observation point. On Fig.\ref{Picture}, the retarded times after each $X_{\text{track}}^{\text{short}}$ step for a given particle are represented by $t_{ret}^1$, $t_{ret}^2$ ,$t_{ret}^3$, $t_{ret}^j$ ... To construct the shower signal at a given observation point $\boldsymbol{x}$ (an antenna position fixed by the user), each particle contribution is added into three histograms corresponding to the three total contributions $\boldsymbol{St}$, $\boldsymbol{C}$ and $\boldsymbol{Cu}$ as a function of reception time $t$.  For these three histograms, the constant time width $\Delta_{\text{size}}^{\text{bin}}$ of bins is fixed by the user (tipically $0.1\text{ ns}<\Delta_{\text{size}}^{\text{bin}}<1\text{ ns}$). For a given particle, to each instant $t_{ret}^j$ is associated a reception time $t^j=t_{ret}^j+R^j(t_{ret}^j)/c$ ; to fill bins of the three histograms, different configurations are possible (we present the method only for the case of $\boldsymbol{Cu}$ histogram but this method is the same to fill $\boldsymbol{St}$ and $\boldsymbol{C}$ histograms) :
\begin{itemize}
\item for bins fully-contained between $t^j=t_{ret}^j+R^j(t_{ret}^j)/c$ and $t^{j+1}=t_{ret}^{j+1}+R^{j+1}(t_{ret}^{j+1})/c$, bins are incremented by :
\begin{equation}
\epsilon*\frac{\boldsymbol{cu}_i(\boldsymbol{x},t^j)+\boldsymbol{cu}_i(\boldsymbol{x},t^{j+1})}{2}
\label{increment}
\end{equation}
with $\epsilon=1$. Eq.\ref{increment} means that in SELFAS2, the electric field emitted by a particle is considered constant over time during a short track ($X_{\text{track}}^{\text{short}}$). This approximation can be made because we consider here that the distance R to the observation point, is large compared to the size of the particle trajectory during a short track (Fraunhofer approximation).
\item if $t^j$ and $t^{j+1}$ are completely contained into a bin time window, the corresponding bin is incremented by Eq.\ref{increment} with a factor $\epsilon$ equal to :
\begin{equation}
\epsilon=\frac{t^{j+1}-t^j}{\Delta_{\text{size}}^{\text{bin}}}
\end{equation}
\item if $t^j$ is contained into a bin $b$, delimited by the time $t_{\text{bin}}^{\text{b}}$ and $t_{\text{bin}}^{\text{b}}+\Delta_{\text{size}}^{\text{bin}}$ and if $t^{j+1}>t_{\text{bin}}^{\text{b}}+\Delta_{\text{size}}^{\text{bin}}$, then, the bin $b$ is incremented by Eq.\ref{increment} with a factor $\epsilon$ equal to :
\begin{equation}
\epsilon=\frac{t_{\text{bin}}^{\text{b}}+\Delta_{\text{size}}^{\text{bin}}-t^j}{\Delta_{\text{size}}^{\text{bin}}}
\end{equation}
\item and finally, if $t^{j+1}$ is contained into a bin $b$, delimited by the time $t_{\text{bin}}^{\text{b}}$ and $t_{\text{bin}}^{\text{b}}+\Delta_{\text{size}}^{\text{bin}}$ and if $t^{j}<t_{\text{bin}}^{\text{b}}$, then the bin $b$ is incremented by Eq.\ref{increment} with a factor $\epsilon$ equal to :
\begin{equation}
\epsilon=\frac{t^{j+1}-t_{\text{bin}}^{\text{b}}}{\Delta_{\text{size}}^{\text{bin}}}
\end{equation}
\end{itemize}
Once all particle trajectories have been considered, the numerical time derivatives for $\boldsymbol{C}(\boldsymbol{x},t)$ and $\boldsymbol{Cu}(\boldsymbol{x},t)$ are performed and the complete electric field corresponding to this observation point is obtained summing up the three total contributions. 

\section{Discussion from a first example: a vertical $10^{17}$ air shower}
To compare the predicted observables from SELFAS2, REAS3 and MGMR (at the end of this section), we simulate a vertical $10^{17}$ eV induced air shower in the Auger site configuration. The geomagnetic field characteristics at the Auger site are $|\boldsymbol{B}|$ = 23 $\mu$T, $\theta_B$ = 58° and $\phi_B$ = 0°  where $\theta_B$ and $\phi_B$ are the zenith angle and the azimuthal angle of the geomagnetic field. The geomagnetic field comes from the south and is oriented upward. The ground shower core of the air shower simulated is located exactly at the center of a dense array composed of 145 antennas. Antenna positions are shown in Fig.\ref{footprint} left. The ground altitude is 1400 m. 

\subsection{Numerical stability}
\begin{figure}
\includegraphics[scale=0.62]{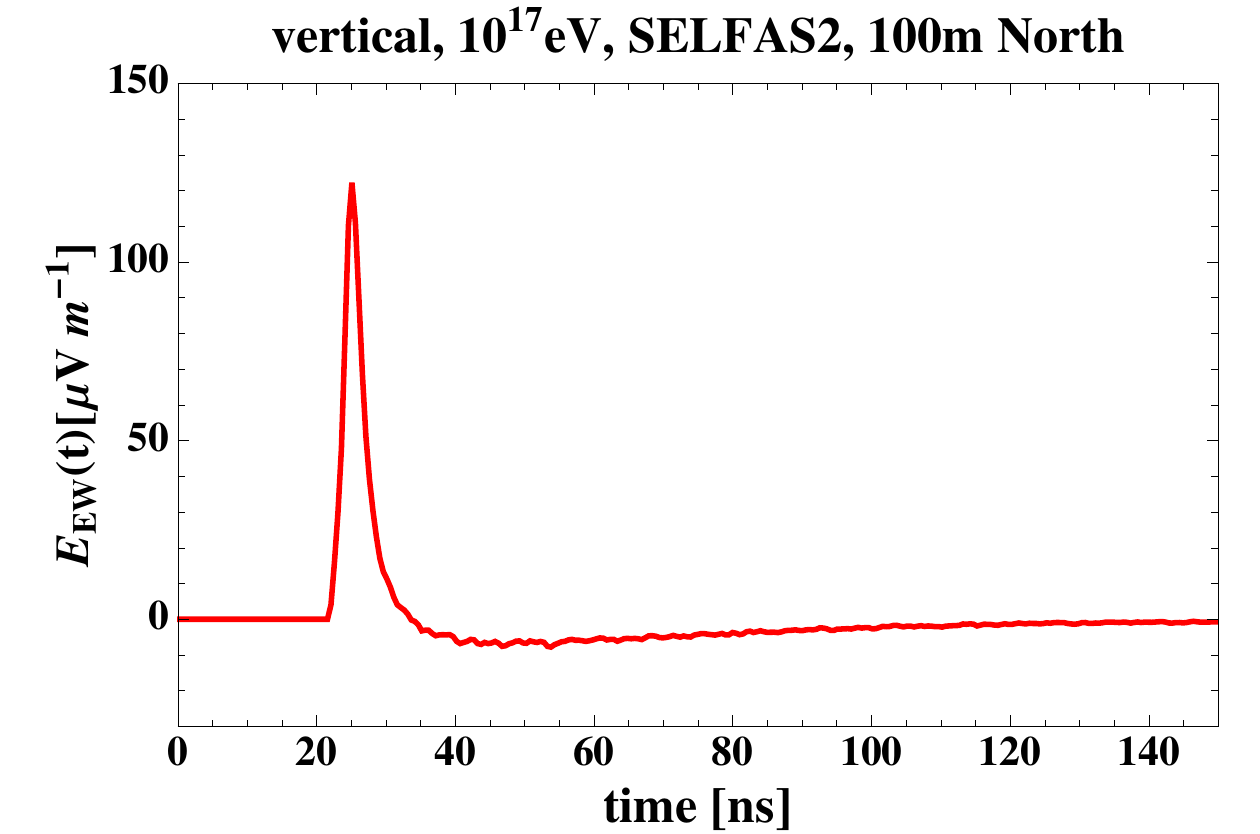}
\includegraphics[scale=0.64]{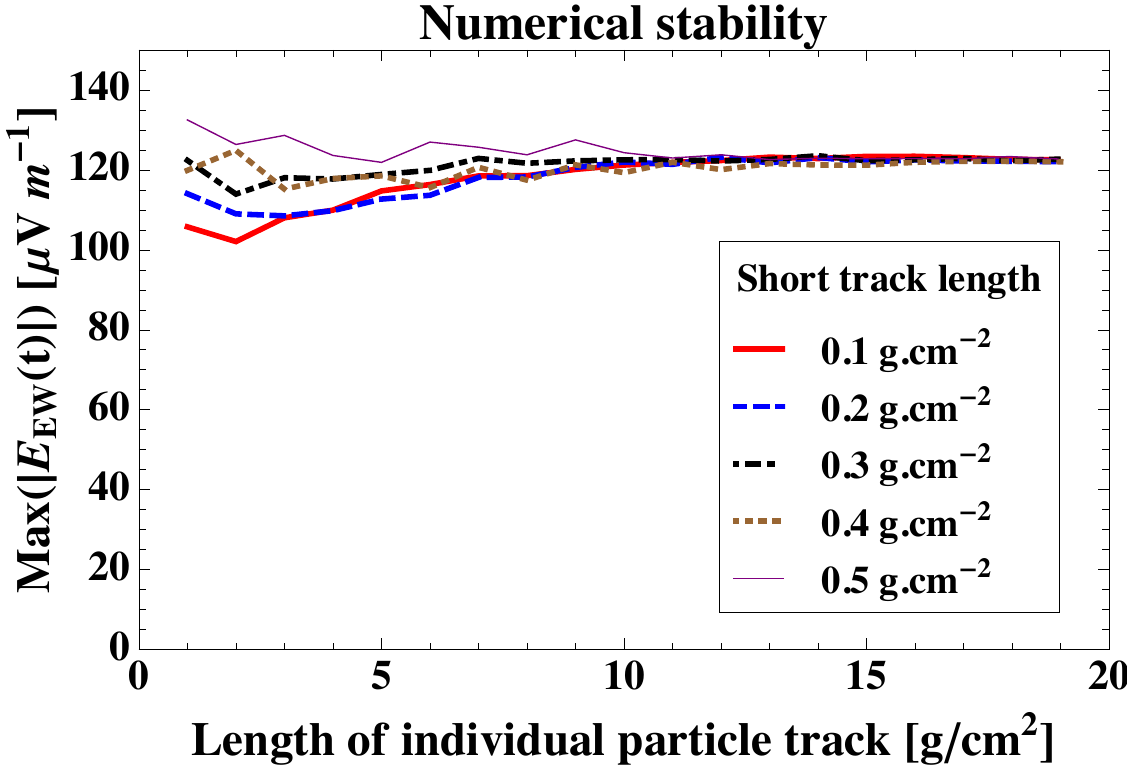}
\caption{\footnotesize{Left: Signal in the EW polarization for an observer located at a distance of 100~m north of the shower core. Right: amplitude of the signal as a function of the length of individual particle track and its resolution (short track length) fixed in SELFAS2.}}
\label{Stability}
\end{figure}

The final result of the observed signal must be independent of the choice of the particle trajectory length discussed in the previous section. With the configuration described above, we show in Fig.\ref{Stability} left the signal for an observer located at a distance of 100~m north of shower core. In Fig.\ref{Stability} we show the amplitude of this signal as a function of the length $l$ of the individual particle track and its resolution (short track length) fixed in SELFAS2. We observe that for small individual track length (smaller than 5~g$\,$cm$^{-2}$), the amplitude of the signal is not stable. In \cite{key27} the angular distributions of particles are given for both electrons and positrons without discern their nature ; to model the effect of the geomagnetic field, electrons and positrons must be propagated. Fig.\ref{Stability} shows that a minimum length for the individual particle track ($l>$10~g$\,$cm$^{-2}$) is necessary to solve the absence of geomagnetic field effect in \cite{key27} and to guarantee the stability of the signal.

The choice of the length of the individual particle track is then made in order to obtain stability but also to stay as close as possible to the parameterized distributions given in \cite{key27}. In SELFAS2 this length is fixed to 15 g.cm$^{-2}$ with a resolution of 0.3 g.cm$^{-2}$.

\begin{figure}
\includegraphics[scale=0.62]{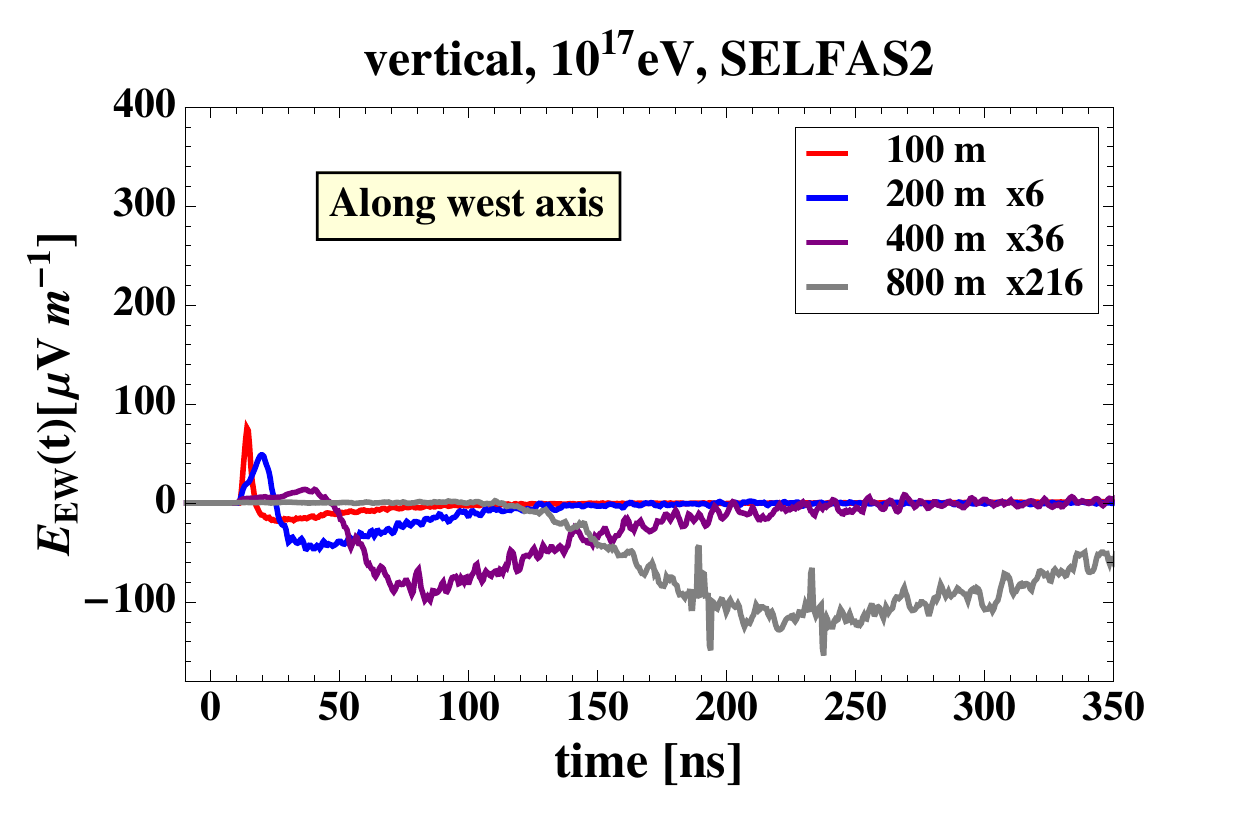}
\includegraphics[scale=0.62]{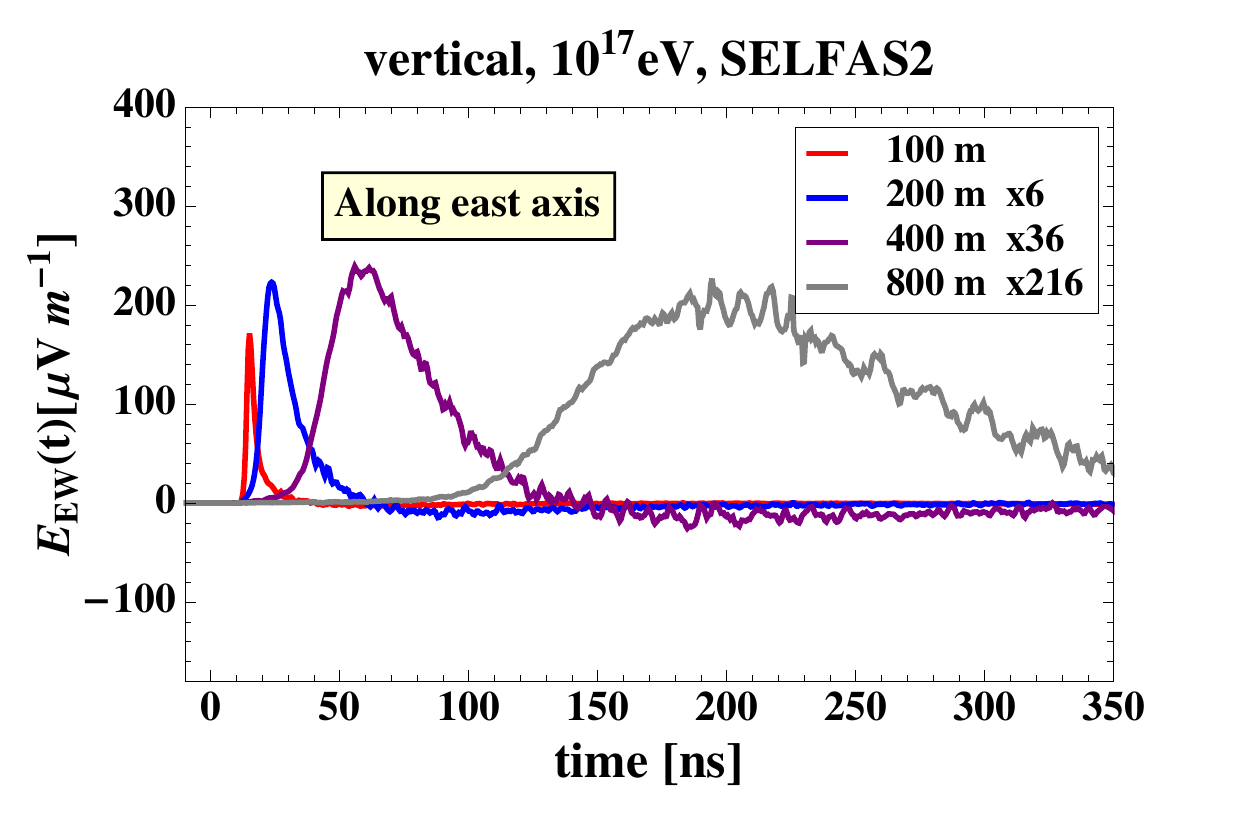}
\includegraphics[scale=0.62]{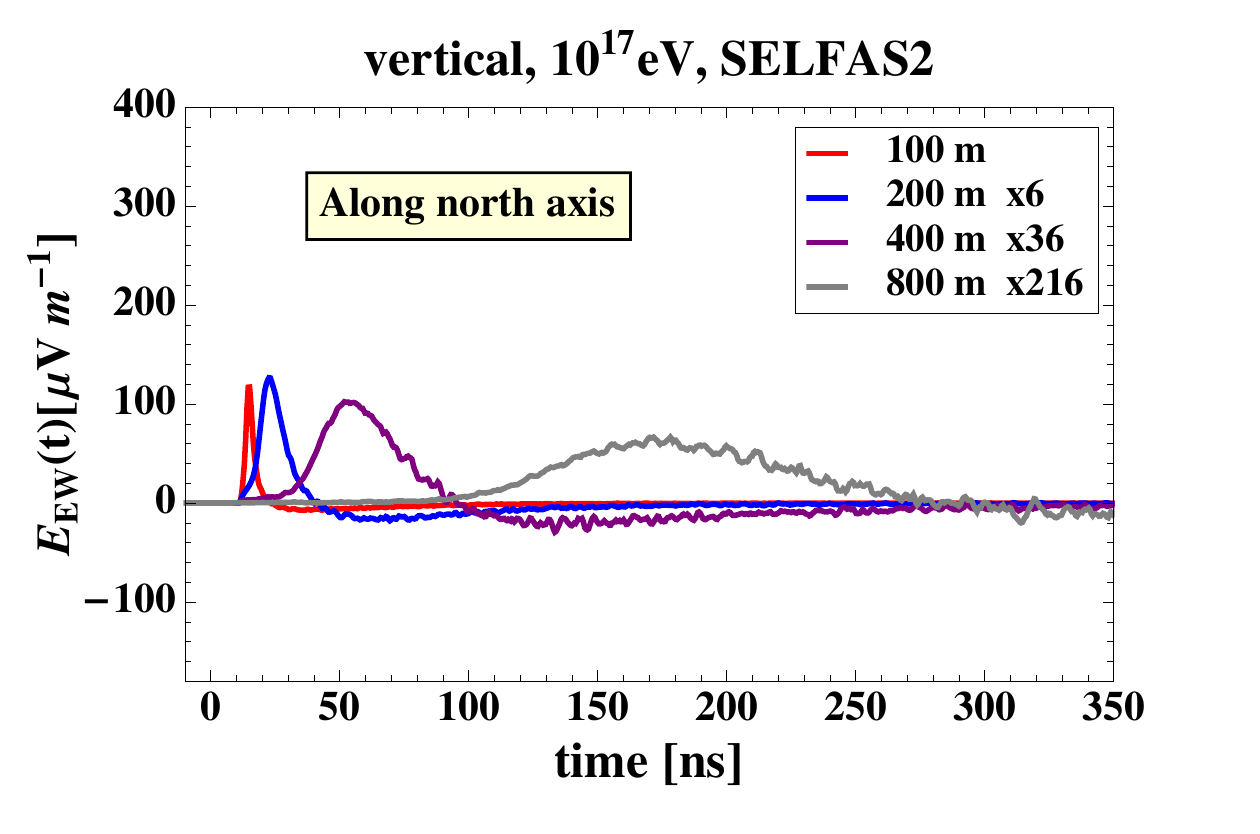}
\includegraphics[scale=0.62]{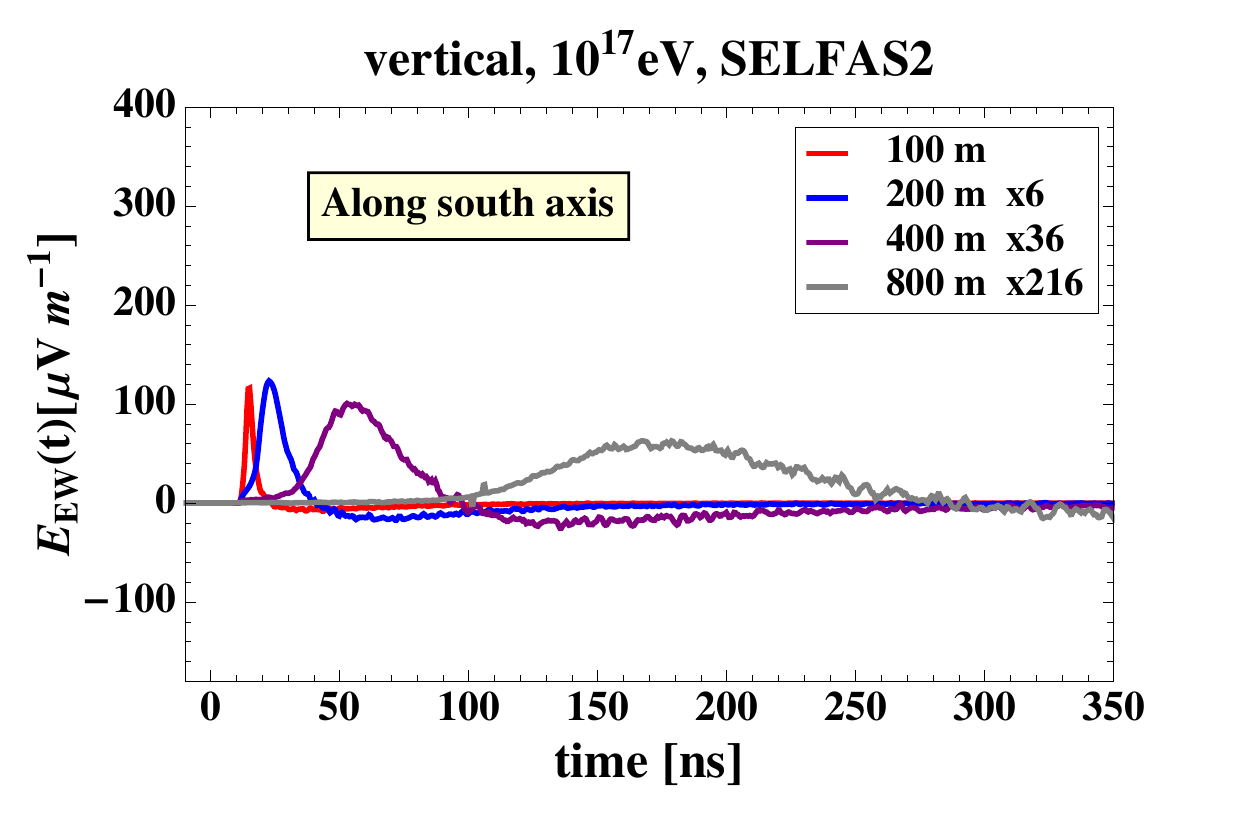}
\caption{\footnotesize{East-west polarization radio pulses for a vertical shower with a primary energy of $10^{17}$ eV computed with SELFAS2. In each figure, we show the results obtained for different distances to shower core (100 m, 200 m, 400 m and 800 m). Top, left: antennas located along the semi-axis starting from the ground shower core pointing toward the west direction. Top, right: antennas located along the semi-axis starting from the ground shower core pointing toward the east direction. Bottom, left: antennas located along the semi-axis starting from the ground shower core pointing toward the north direction. Bottom, right: antennas located along the semi-axis starting from the ground shower core toward pointing the south direction.}}
\label{ImpulsionsAuger}
\end{figure}

\subsection{Results}
\begin{figure}
\begin{center}
\includegraphics[scale=0.62]{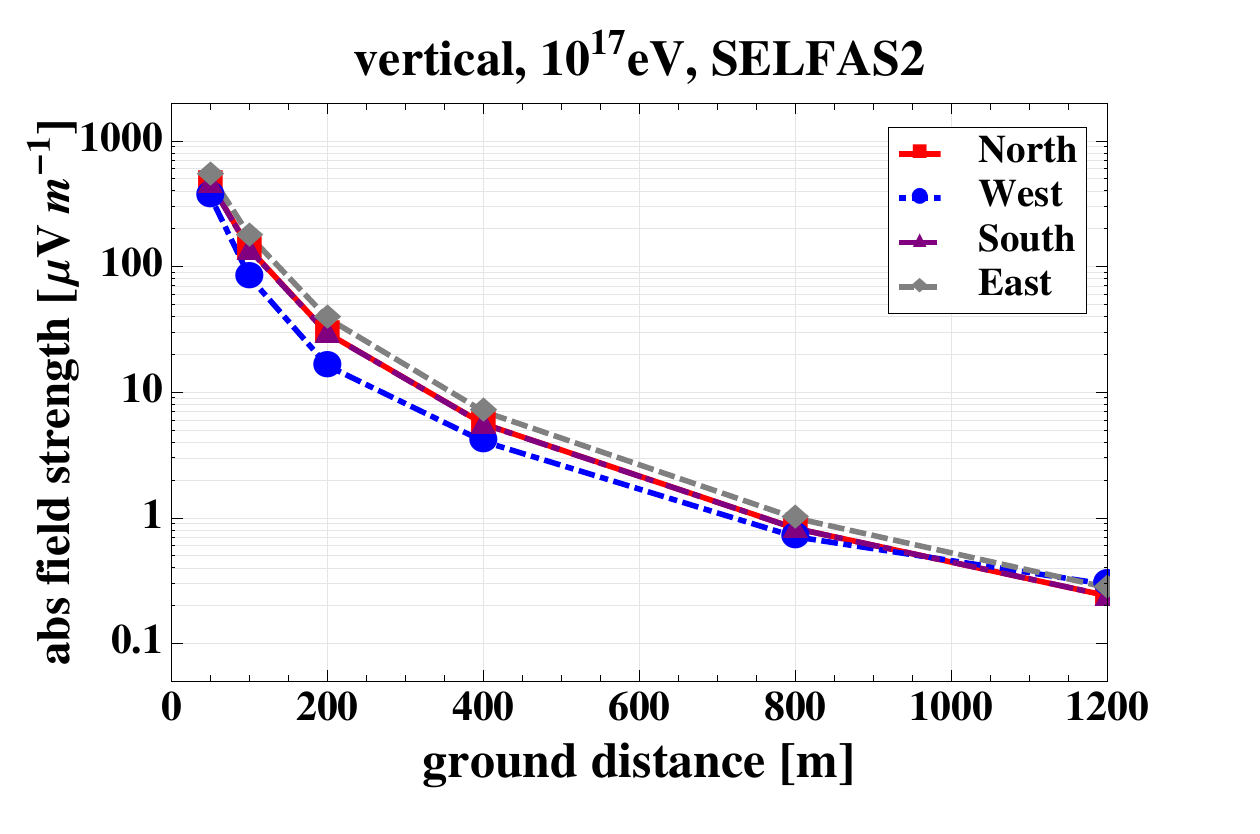}
\includegraphics[scale=0.5]{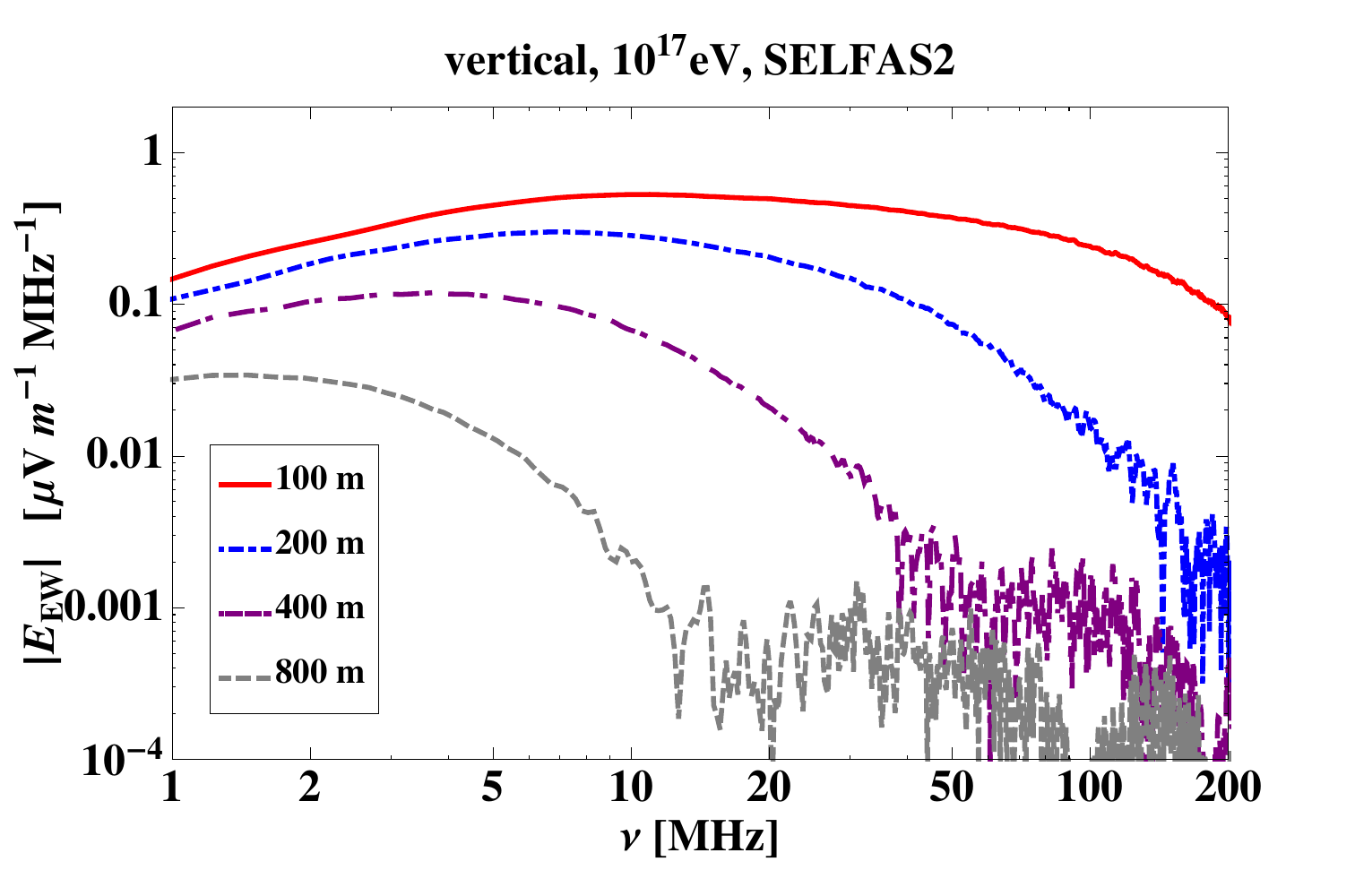}
\caption{\footnotesize{Left: lateral distribution of the absolute field strength pulse amplitude for a $10^{17}$ eV vertical air shower. Right: frequency spectra of a $10^{17}$ eV vertical air shower seen from different distances to the shower core. East-west polarization.}}
\label{ProfilesSpectras}
\end{center}
\end{figure}

In Fig.\ref{ImpulsionsAuger} we show the east-west polarization radio pulses in the time domain for a vertical shower with a primary energy of $10^{17}$ eV computed with SELFAS2. The ground frame is arranged so that the air shower core is located at $x=0$ and $y=0$. The four figures represent the pulses observed by antennas set along axis oriented toward the west, the east, the north and the south. In these figures we clearly see that the decay of the pulse amplitude along the axis west-east is not symmetric around the air shower axis. This effect is more visible in Fig.\ref{ProfilesSpectras} left where we show the lateral distribution of the absolute field strength pulse amplitude. This asymmetry directly comes from the influence of the net charge excess (more details in the next subsection). The spectral composition of the pulses observed in Fig.\ref{ImpulsionsAuger} is obtained by Fourier analysis. The result is shown in Fig.\ref{ProfilesSpectras} right. We see in this figure that the frequency where the coherence is lost, decreases with distance to the shower axis. 

\begin{figure}[H]
\begin{center}
\includegraphics[scale=0.52]{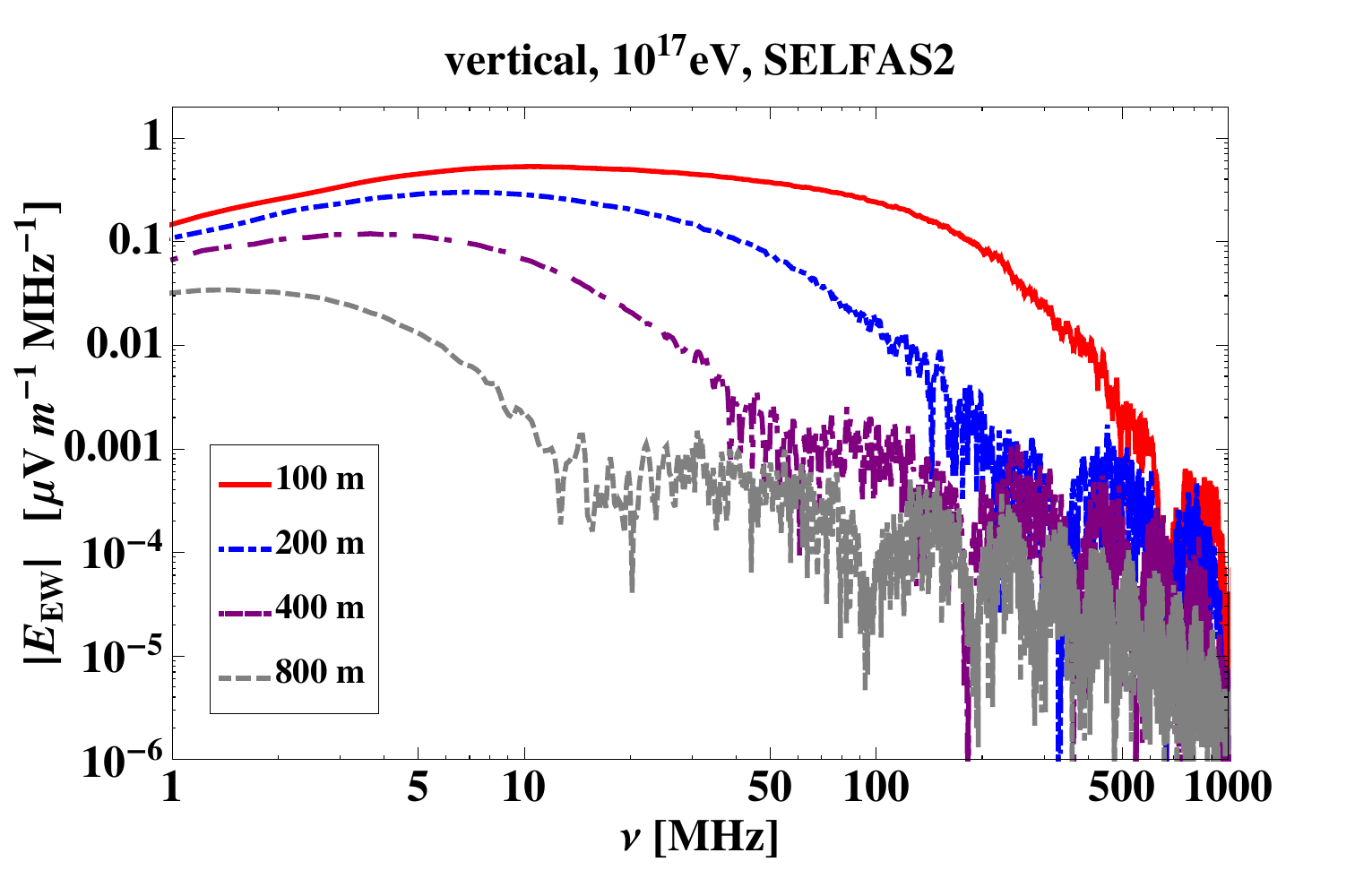}
\includegraphics[scale=0.5]{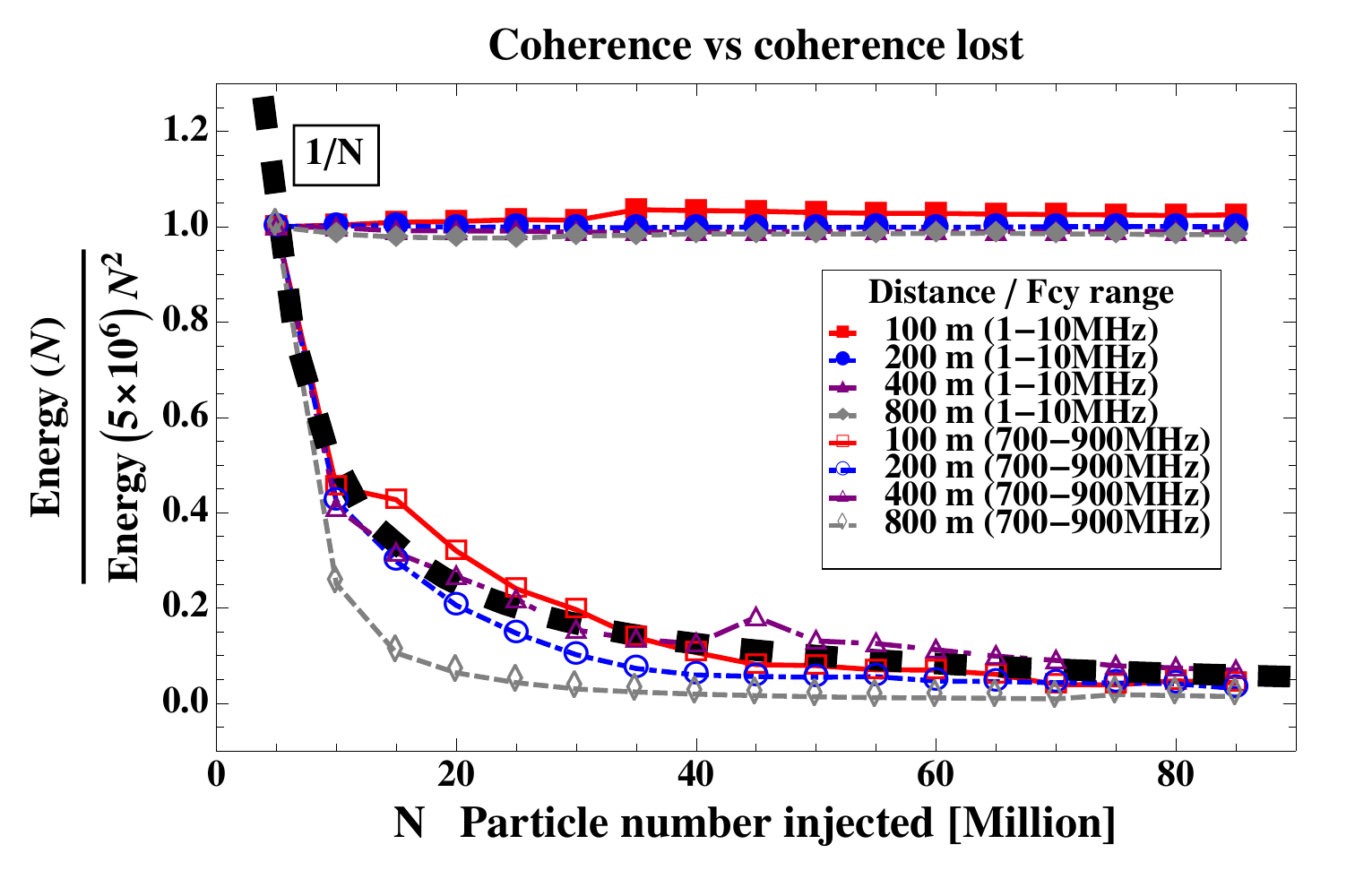}
\caption{\footnotesize{Left: Spectra used for the calculation of the energy deposited as a function of the number of particles injected in the simulation. The integrations of spectra are performed every five million particles injected, for the two frequency ranges 1-10 MHz and 700-900 MHz. Right: energy deposited as a function of number of particles, for different distances to air shower axis. The difference of behavior between the two frequency ranges of integration is due to the coherence loss of the signal from a particular frequency which decreases with distance to air shower axis.}}
\label{Integral}
\end{center}
\end{figure}

To illustrate the notion of coherence loss, let's compute the evolution of energy deposited by the air shower as a function of the number of particles injected in the simulation. Every five million particles injected, we performed the spectra integrations in the range 1-10 MHz and in the range 700-900 MHz (see Fig.\ref{Integral} left). In Fig.\ref{Integral} right, we show the energy deposited behavior as a function of the number of particles for the two frequency ranges and for different distances to air shower axis. We see that the energy deposited in the low frequency range depends on the square of particles number whereas the energy deposited in the range 700-900 MHz depends on the number of particles (see comparison with $1/N$ in Fig.\ref{Integral} right). This phenomena, due to the coherence loss, is naturally taken into consideration in SELFAS2.
\subsection{Transverse current vs charge excess}
In section 2, we showed that the EAS radio emission is the summation of three terms. The static contribution, $\boldsymbol{St}(\boldsymbol{x},t)$, is approximatively two orders of magnitude lower than the two other contributions, so in the following, this contribution will be not discussed. Finally, the time derivative of the transverse current and the time derivative of the net charge excess are in competition. In most cases, the transverse current contribution is dominant with respect to the net charge excess contribution. This is not true for air showers parallel to the 
\begin{figure}[h]
\includegraphics[scale=0.45]{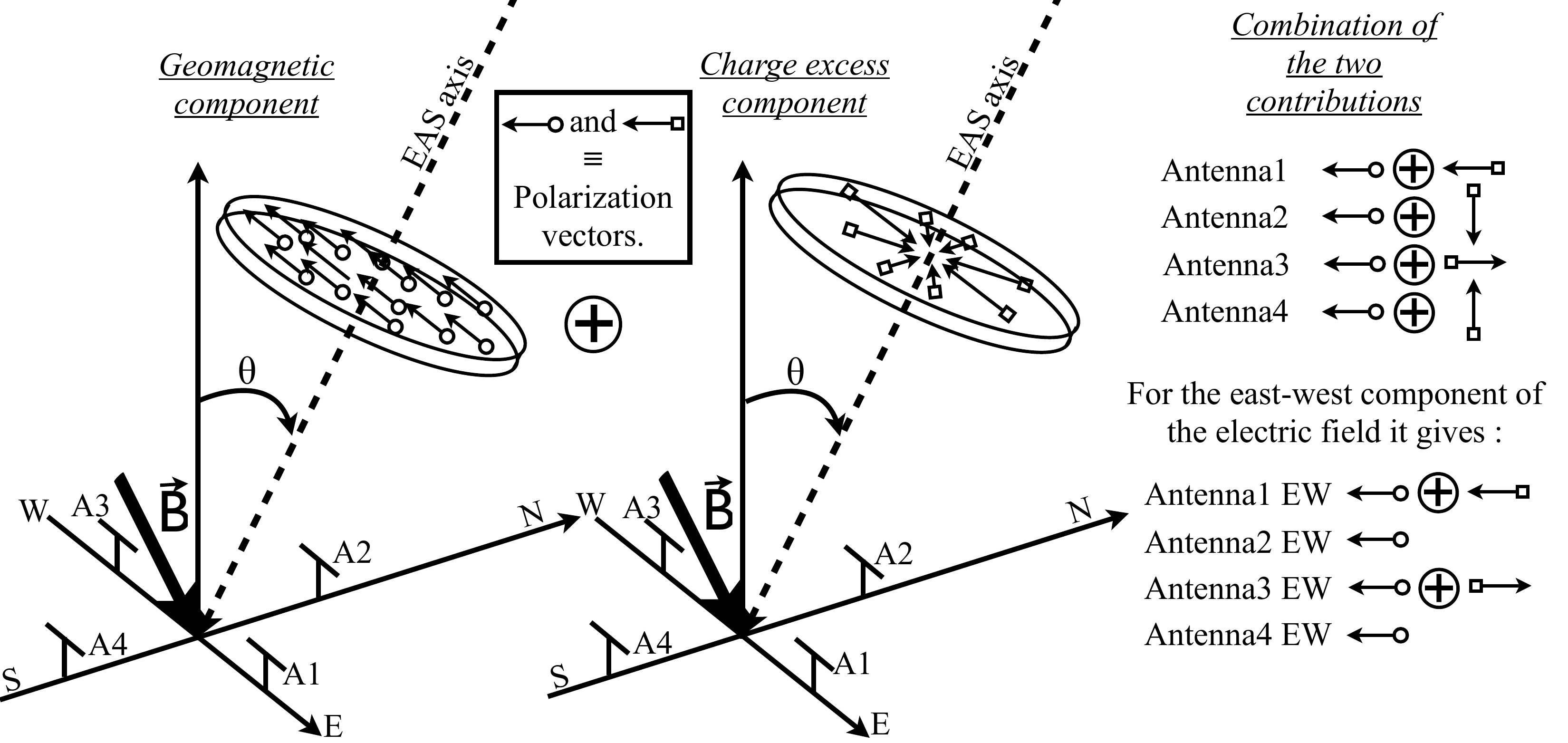}
\caption{\footnotesize{Polarization vectors of the transverse current and the net charge excess contributions in the plane perpendicular to the shower axis. Due to the fact that the polarization vectors of these two contributions are not always oriented in the same direction, their combination can be constructive or destructive in function of the antenna position. For the east-west polarization of the electric field (as it is given in Fig.\ref{ImpulsionsAuger} and Fig.\ref{ProfilesSpectras} left), the signal amplitudes observed by antennas located on the east side of the plane containing the ground shower core and the geomagnetic field appear finally higher than the signal amplitudes of antennas located on the other side.}}
\label{Picture2}
\end{figure}
\begin{figure}[!h]
\hspace{-1.cm}\includegraphics[scale=0.55]{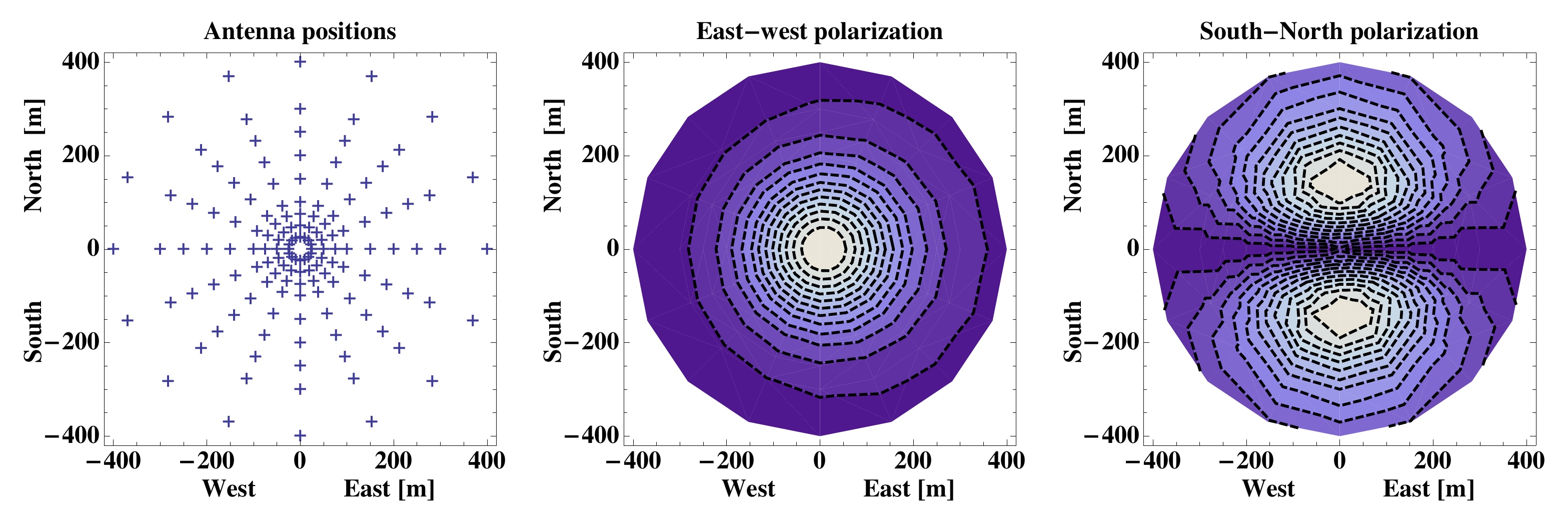}
\caption{\footnotesize{Left: Antennas positions used for ground footprint calculation. Center and right: Ground footprint of the energy deposited by a vertical $10^{17}$ eV air shower below 85 MHz, for the two horizontal polarizations. The east-west polarization shows a slight east-west asymmetry which is due to the net charge excess contribution (see text for more details).}}
\label{footprint}
\end{figure}
geomagnetic field. In such cases, systematic deviations of particles due to the geomagnetic field disappear and the transverse current contribution vanishes. To understand the behavior of the two dominant contributions combination, we show in Fig.\ref{Picture2} the polarization vectors of the transverse current and the net charge excess contributions in the plane perpendicular to the shower axis. Due to the fact that the polarization vectors of these two contributions are not always oriented in the same direction, their combination can be constructive or destructive. In the configuration of the Auger site for instance, the summation is then constructive for the case of antennas located on the east side of the plane containing the ground shower core and the geomagnetic field. On the west side of this plane the summation is destructive. This asymmetry effect is also observable in an other form in Fig.\ref{footprint} where we present the ground footprint of the two horizontal polarizations. Fig.\ref{footprint} are obtained using a dense array composed of 145 antennas (cf Fig.\ref{footprint} left) for which we calculate the energy deposited below 85 MHz, for the two horizontal polarizations. The results obtained here confirm results previously obtained and discussed in \cite{key16,key02}.

\begin{figure}[h]
\begin{center}
\hspace{-0.1cm}\includegraphics[scale=0.1145]{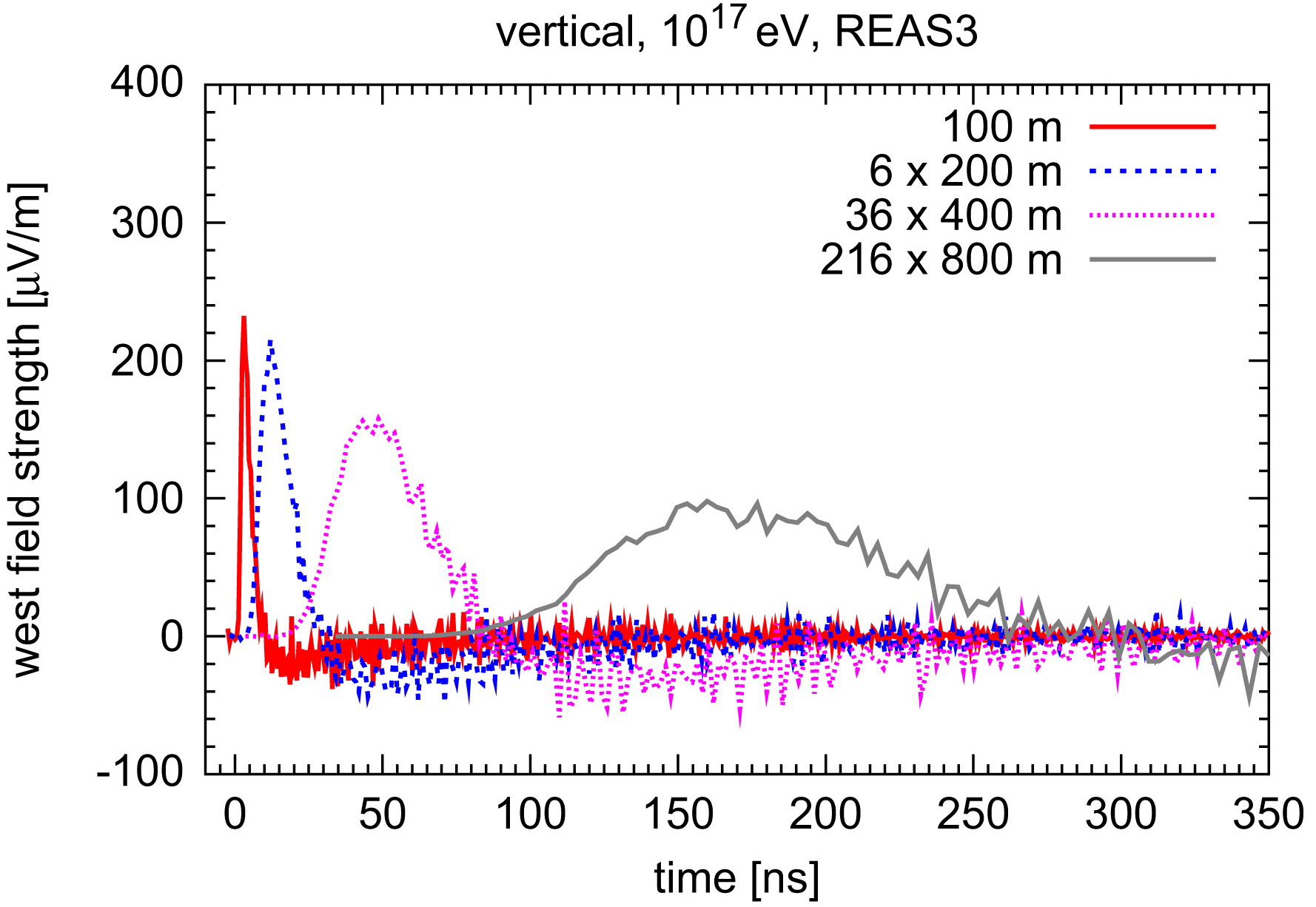}\hspace{0.75cm}
\includegraphics[scale=0.1145]{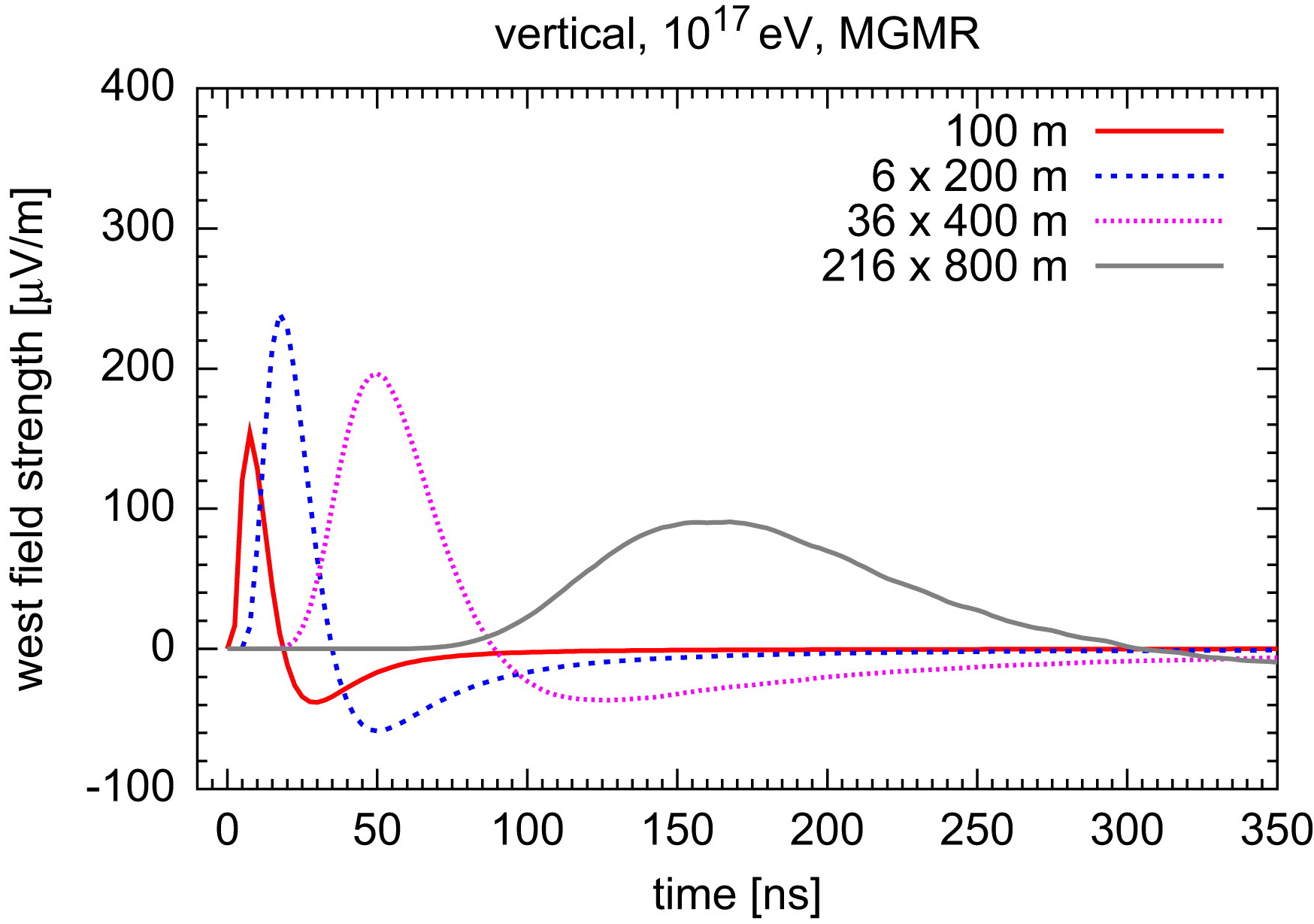}
\includegraphics[scale=0.1145]{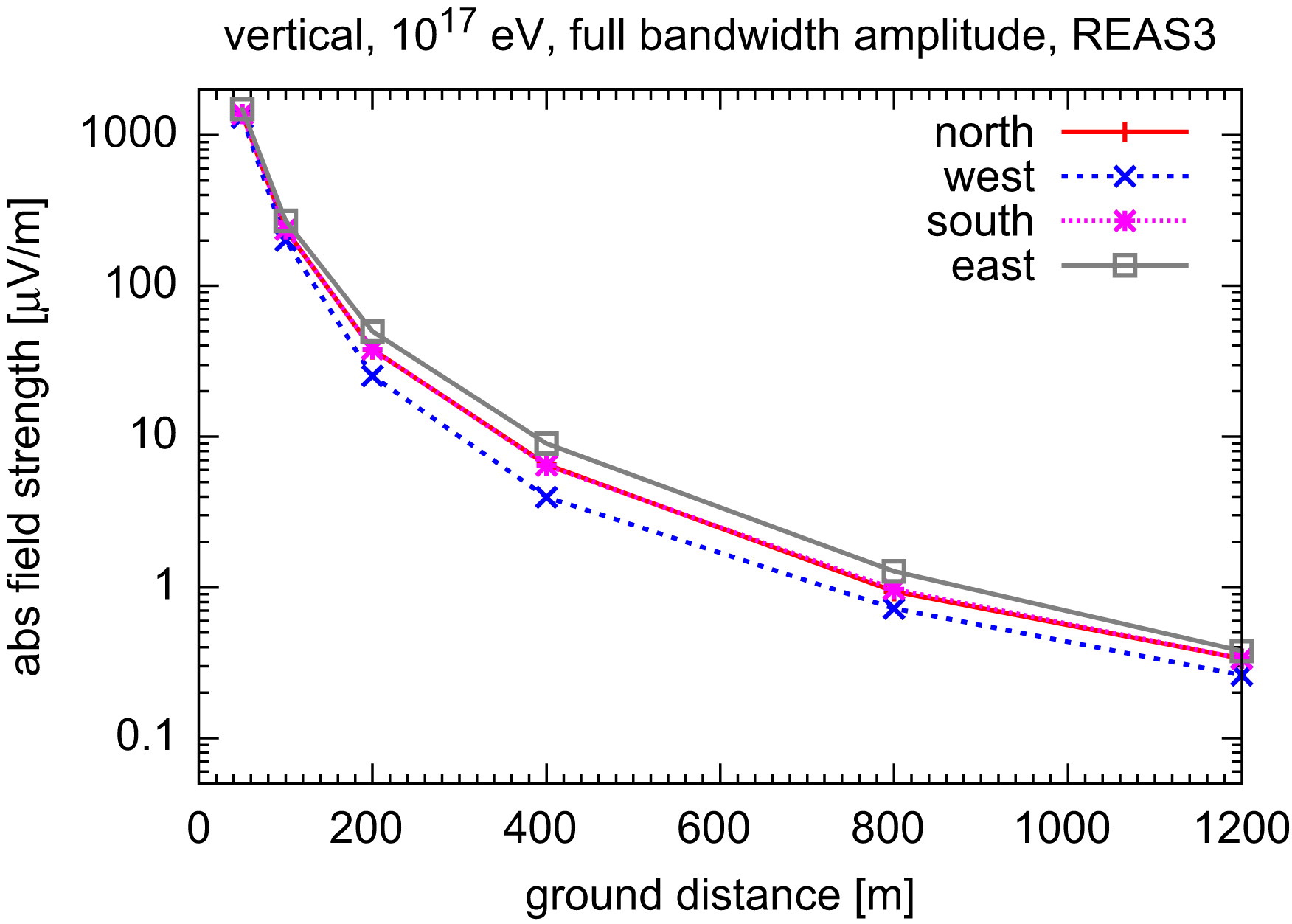}\hspace{0.8cm}
\includegraphics[scale=0.1145]{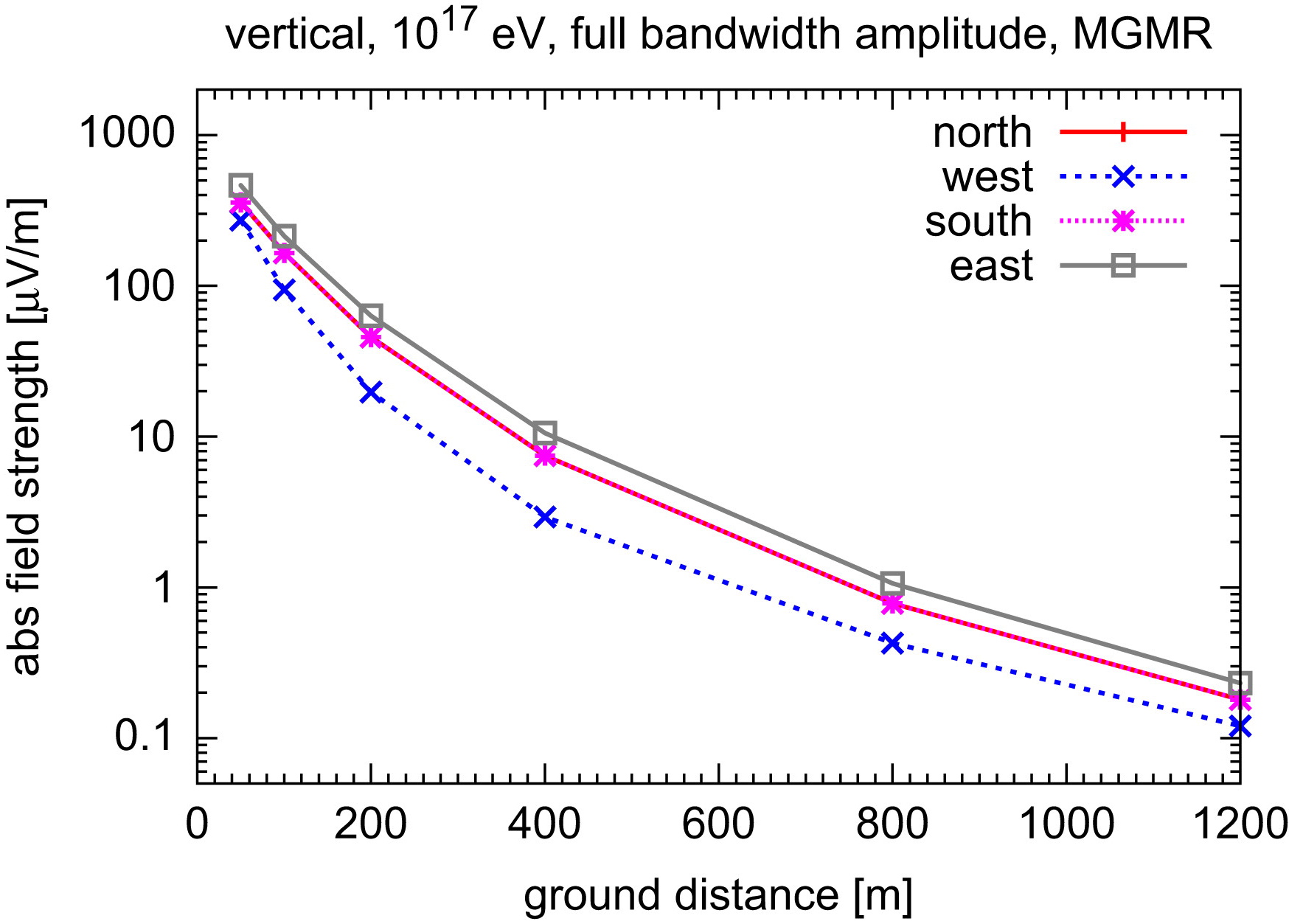}
\caption{\footnotesize{From \cite{key19}. Top: Comparison of the western polarization unlimited bandwidth radio pulses for vertical showers with a primary energy of $10^{17}$ eV for REAS3 (left) and MGMR (right). Bottom: Comparison of the absolute field strength unlimited bandwidth pulse amplitude lateral distribution for vertical showers with a primary energy of $10^{17}$ eV for REAS3 (left) and MGMR (right). These figures can be directly compared with Fig.\ref{ImpulsionsAuger} and Fig.\ref{ProfilesSpectras}.}}
\label{REAS3MGMR}
\end{center}
\end{figure}
\subsection{Comparison between SELFAS2, REAS3 and MGMR}
In a recent paper \cite{key19}, a direct comparison is done between the two main approaches discussed in the section 2. They show in this paper that after great disagreements, the two models finally give results which converge in the same direction. We extracted some figures of this paper to make a direct comparison with the results obtained with SELFAS2. In Fig.\ref{REAS3MGMR} we show the results obtained by REAS3 and MGMR for the same air shower simulated with SELFAS2 in subsection 4.1. These figures can be directly compared with Fig.\ref{ImpulsionsAuger} and Fig.\ref{ProfilesSpectras}. The agreement between the three models is satisfactory but it still remains some differences which can be due to the electric field formalism adopted and to the shower description. For a better comparison, the simulations performed with REAS3 and MGMR are based on EAS characteristics coming from the same CORSIKA shower output file. SELFAS2 doesn't use any simulated shower, but only the analytical geometrical distributions previously described (based on CORSIKA simulation). The agreement between the three models shows that their implementation in SELFAS2 gives promising results.

\section{Conclusion}

With SELFAS2, we successfully managed to construct a complete autonomous tool able to help detailed understanding in the mechanisms involved in the air showers radio emission. The concept of universality of air showers feature permits to avoid the heavy use of air shower generators and to obtain acceptable results after few minutes for one antenna simulated. Furthermore, we confirmed that EAS radio emission is mainly due to two different contributions: the time variation of the transverse current (corresponding to the third term in Eq.\ref{SumField}) and the time variation of the EAS macroscopic charge excess (corresponding to the second term in Eq.\ref{SumField}). The polarization patterns of these contributions are different leading to asymmetries in both the EAS ground footprint and the lateral profile. It would be very interesting to search for such asymmetries in the data (CODALEMA, LOPES, RAuger) in order to make progress in the understanding of the radio emission. The quantification of the experimental radio asymmetry should be a good estimator of the electrons-positrons asymmetry. 

Today, we can clearly say that air shower radio simulations converge in the same direction, showing good agreements even using different approaches. Coupled analysis between data and SELFAS2 are currently in progress to test the validity of theoretical predictions.

\end{document}